\documentclass[aps,amsmath,onecolumn,amssymb,nofootinbib,floatfix,showpacs]{revtex4}
\usepackage{graphicx}
\usepackage{epsfig}
\usepackage{subfigure}
\usepackage{graphics,color}
\usepackage[none]{hyphenat}
\usepackage[dvipdfmx]{hyperref}
\usepackage{amsbsy}
\usepackage{bm}
\newcommand{\be}{\begin{equation}}
\newcommand{\ee}{\end{equation}}
\newcommand{\bea}{\begin{eqnarray}}
\newcommand{\eea}{\end{eqnarray}}
\newcommand{\nn}{\nonumber}
\newcommand{\del}{\partial}
\newcommand{\Tr}{{\mathrm{Tr}}}

\begin{document}

\title{Revisiting the Phase Structure of the Polyakov-quark-meson Model in the presence of Vacuum Fermion Fluctuation.}
\author{Uma Shankar Gupta}
\email {guptausg@gmail.com}
\author{Vivek Kumar Tiwari}
\email {vivekkrt@gmail.com}
\affiliation{Department of Physics, University of Allahabad, Allahabad 211002, India.}

\date{\today}
\begin{abstract}
             We have  considered the contribution of fermionic vacuum loop in the effective potential
of Polyakov loop extended Quark Meson Model (PQM) for the two quark flavour case and explored 
the phase structure and thermodynamics of the resulting PQMVT model (Polyakov Quark Meson Model 
with Vacuum Term) in detail at non zero as well as zero chemical potential.
The temperature variations of order parameters and their derivatives have been calculated
and the phase diagram together with the location of critical end point (CEP) has been obtained in $\mu$, 
and $T$ plane in both the models PQMVT and PQM. The PQMVT model analysis has been compared with the 
calculations in PQM model in order to bring out the effect of fermionic vacuum term on the physical observables. 
We notice that the critical end point (CEP) which is located near the temperature axis
at ($\mu=81.0$, $T = 167$ MeV) in the PQM model gets shifted close to the chemical potential axis at 
($\mu_{CEP}=294.7$, $T_{CEP}=84.0$ MeV) in the PQMVT model calculations of the phase diagram. Since it emerges from a  
background of second order  transition in the  chiral limit of massless quarks, the  crossover occurring at  
$\mu$ = 0 in PQMVT model for the realistic case of explicitly broken chiral symmetry, has been identified as  
quite a soft and smooth transition. We have presented  and compared the results for temperature variations of thermodynamic observables at zero and different non-zero quark chemical potentials. It is noticed that the presence of fermionic vacuum 
term in the effective potential leads to a smoother and slower temperature variation of thermodynamic quantities.

\end{abstract}


\maketitle

\section{Introduction}
\label{sec:intr}

Quantum Chromo-dynamics (QCD), the commonly accepted theory of strong interaction  
predicts that normal hadronic matter undergoes a phase transition, 
where the individual hadrons dissolve into their constituents and 
produce a collective form of matter known as the Quark Gluon Plasma 
(QGP) under the extreme conditions of high temperature and/or density
\cite{Rischke:03,Ortmanns:96ea,Muller,Svetitsky}. Relativistic heavy ion collision 
experiments at RHIC (BNL), LHC (CERN) and the future CBM experiments 
at the FAIR facility (GSI-Darmstadt) aim to create and study such a 
collective state of matter. Study of the different aspects of this 
phase transition, is a tough and challenging task because Quantum 
Chromodynamics (QCD) becomes nonperturbative in the low energy limit.
 
It is  well known that the basic QCD Lagrangian has the global
$SU_{L+R}(N_{f}) \times SU_{L-R}(N_{f})$ symmetry for $ N_{f} $ flavours 
of massless quarks. The axial (A=L+R) part of this symmetry known as the
chiral symmetry is spontaneously broken by the formation of a chiral condensate
in the low energy hadronic vacuum of QCD and one gets $ (N_{f}^2-1) $ massless
Goldstone bosons according to the Goldstone's theorem. Since quarks are not 
massless in real life, chiral symmetry of the QCD lagrangian gets explicitly broken 
and massless modes become pseudo-Goldstone bosons after acquiring mass. Nevertheless,
the observed lightness of pions in nature suggests that we have an approximate 
chiral symmetry for QCD with two falvours of light u and d quarks.
In the opposite limit of infinitely heavy quarks, QCD becomes a pure $SU(N_{c})$
gauge theory which remains invariant under the global $Z(N_{c})$ center symmetry of
the color gauge group. The Center symmetry which is a symmetry of hadronic vacuum, 
gets spontaneously broken in the high temperature/density regime of QGP. 
The expectation value of the Wilson line (Polyakov loop) is related 
to the free energy of a static color charge. It vanishes in the confining phase as 
the quark has infinite free energy and becomes finite in the deconfined phase.
Hence the Polyakov loop serves as the order parameter of the confinement-deconfinement 
phase transition \cite{Polyakov:78plb}. Even though the center symmetry is always broken 
with the inclusion of dynamical quarks in the system, one can regard the 
Polyakov loop as an approximate order parameter because it is a good 
indicator of a rapid crossover in the confinement-deconfinement transition \cite{Pisarski:00prd,Vkt:06}.
 
The first principle lattice QCD Monte Carlo simulations
(see e.g.~\cite{Karsch:02,Fodor:03, AliKhan:2001ek, Allton:05,Aoki:06,Karsch:05,
Karsch:07ax,Cheng:06,Cheng:08,Digal:01}) give us important information and insights 
regarding various aspects of the QGP transition,
like the restoration of chiral symmetry in QCD, order of the 
confinement-deconfinement phase transition, richness of the QCD phase 
structure and mapping of the phase diagram. Unfortunately progress in lattice QCD 
calculations has got severely hampered due to the QCD action becoming complex on account 
of the fermion sign problem \cite{Karsch:02} when baryon density/chemical potential is non zero. 
Though several methods have been developed to circumvent the sign problem 
at small baryon chemical potentials, a general solution to the 
sign problem for all chemical potentials is yet to be devised. Further since lattice 
calculations are technically involved and various issues are not conclusively settled 
within the lattice community, one resorts to the calculations within the
ambit of phenomenological models developed in terms of effective degrees 
of freedom. These models serve to complement the lattice simulations and 
give much needed insight about the regions of phase diagram inaccessible 
to lattice simulations.

In recent years, effective chiral models, having the pattern of chiral symmetry breaking 
as that of QCD like the linear sigma models(LSM) \cite{Rischke:00,Hatsuda,Chiku,Herpay:05,
Herpay:06,Herpay:07,Fejos},the quark-meson (QM) models(see e.g.\cite{Mocsy:01prc, Schaefer:09,
Andersen,jakobi,mocsy,bj,Schaefer:2006ds,Kovacs:2006ym,Bowman:2008kc,Jakovac:2010uy,koch}), 
 Nambu-Jona-Lasinio (NJL) type models \cite{Costa:08, Mocsy:01prc, Kneur,kahara,nickel}, 
have led to the investigation of the properties and structure of chiral symmetry restoring phase 
transition at sufficiently high temperature and density. Further these models were extended
to incorporate the features of confinement-deconfinement transition where chiral condensate and Polyakov 
loop got simultaneously coupled to the quark degrees of freedom. Thus Polyakov loop augmented 
PNJL models \cite{Fukushima:04plb,Ratti:06, Ratti:07,Ratti:07npa,Tamal:06,Sasaki:07,Hell:08,
Abuki:08,Ciminale:07,Fu:07,Fukushima:08d77,Fukushima:08d78,Fukushima:09,Ratti:07prd,Costa:09,mats,nonlocal}
,PLSM models and PQM models\cite{Schaefer:07,Schaefer:08ax,Schaefer:09wspax,
Schaefer:09ax,H.mao09,gupta,Marko:2010cd,Skokov:2010sf,Herbst:2010rf}
have facilitated the investigation of the full QCD thermodynamics and phase structure at 
zero and finite quark chemical potential and it has been shown that bulk thermodynamics of the effective 
models agrees well with the lattice QCD data.

In most of the QM/PQM model calculations, the fermion
vacuum contributions to the free energy is frequently neglected\cite{Mocsy:01prc,Schaefer:2006ds,
Schaefer:09,kahara,Bowman:2008kc} because
here, the spontaneous breaking of chiral symmetry is generated by the mesonic potential itself. 
While in the NJL/PNJL model investigations,fermion vacuum term leads to the 
dynamical breaking of the chiral symmetry, hence it gets explicitly included up to 
a momentum cutoff $\Lambda$. 
Very recently, it has been shown by Skokov et. al. in Ref. \cite{Skokov:2010sf}
that in a mean field approximation, where the fermion vacuum contribution to the free energy is neglected, 
the order of the phase transition for two flavour QM model in the massless chiral limit becomes first order 
at zero baryon chemical potential. They have further shown that the quark-meson model, with  
appropriately renormalized fermionic vacuum fluctuations in the thermodynamic potential,
becomes an effective QCD-like model because now it can reproduce the second order chiral phase transition at $\mu=0$
as expected from the universality arguments\cite{Wilczek} for the two massless flavours of QCD.
It has also been shown that in the presence of an external  magnetic field, the structure of the phase 
diagram in the PQM model is considerably affected by the fermionic vacuum contribution~\cite{Mizher:2010zb}.
In this paper, we will investigate the effect of fermionic vacuum fluctuations on the phase structure and  
thermodynamics of PQM/QM models in detail at non zero as well as zero chemical potential. In order to bring out
the effect of fermionic vacuum term on the physical observables, we will
compare the results of our calculation with the corresponding PQM model calculations without vacuum term.  
 
The arrangement of this paper is as follows. In Sec.\ref{sec:model},
we have given the formulation of  PQM model for the two quark flavour. 
The Polyakov loop potential and the thermodynamic grand potential 
has been given in subsection \ref{subsec:Plgtp}.
After giving a brief description of the appropriate renormalization of fermionic vacuum 
loop contribution, the subsection \ref{subsec:Vterm} describes how the new  model parameters
are obtained in vacuum when renormalized vacuum term is added to the effective potential.
The section \ref{sec:psvt} investigates the effect of fermionic vacuum term on the 
phase structure and thermodynamics. The subsection \ref{subsec:Phastruct}
explores how, the temperature variation of order parameters and their derivatives at different chemical potentials, 
structure of the phase diagram in the $\mu$ and T plane and the location of critical end point,
gets affected in the presence of vacuum term. The effect on the temperature variation of  thermodynamic observables
namely pressure, entropy, energy density  and interaction measure has been discussed in the subsection \ref{subsec:Topeed}
while the discussion of specific heat, speed of sound and $\frac{p(T)}{\epsilon (T)}$ has been presented
in subsection \ref{subsec:sheat} and finally
the subsection \ref{subsec:Qnqs} describes the results for quark number density and quark number susceptibility.
Summary together with the conclusion has been presented in Sec. \ref{sec:smry}.
The first and second partial derivatives of $\cal U_{\text{log}}$ and $\Omega_{\mathrm{q\bar{q}}}^{\rm T}$ 
with respect to temperature and chemical potential has been evaluated in appendix A.

\section{Model Formulation}
\label{sec:model}

We will be working in the two flavor quark meson linear 
sigma model which has been combined with the Polyakov loop potential
\cite{Schaefer:07}
In this model, quarks coming in two flavor are coupled to the 
$SU_L(2)\times SU_R(2)$ symmetric four mesonic fields $\sigma$ and $\vec\pi$ 
together with spatially constant temporal gauge field represented by Polyakov loop 
potential. Polyakov loop field $\Phi(\vec{x})$ is defined as the 
thermal expectation value of color trace of Wilson loop in temporal 
direction 
\be
\Phi = \frac{1}{N_c}\Tr_c L, \qquad \qquad  \Phi^* = \frac{1}{N_c}\Tr_c L^{\dagger}
\ee

where L(x) is a matrix in the fundamental representation of the 
$SU_c(3)$ color gauge group.
\be
\label{eq:Ploop}
L(\vec{x})=\mathcal{P}\mathrm{exp}\left[i\int_0^{\beta}d \tau
A_0(\vec{x},\tau)\right]
\ee
Here $\mathcal{P}$ is path ordering,  $A_0$ is the temporal component
of Euclidean vector field and $\beta = T^{-1}$ \cite{Polyakov:78plb}.
 
The model Lagrangian is written in terms of quarks, mesons, couplings 
and Polyakov loop potential ${\cal U} \left( \Phi, \Phi^*, T \right)$.

\be
\label{eq:Lag}
{\cal L}_{PQM} = {\cal L}_{QM} - {\cal U} \big( \Phi , \Phi^* , T \big) 
\ee
where the Lagrangian in quark meson linear sigma model
\bea
\label{eq:Lqm}
{\cal L}_{QM} = \bar{q_f} \, \left[i \gamma^\mu D_\mu - g (\sigma + i \gamma_5
  \vec \tau \cdot \vec \pi )\right]\,q_f + {\cal L}_{m}
\eea
The coupling of quarks with the uniform temporal background gauge 
field is effected by the following replacement 
$D_{\mu} = \partial_{\mu} -i A_{\mu}$ 
and  $A_{\mu} = \delta_{\mu 0} A_0$ (Polyakov gauge), where 
$A_{\mu} = g_s A^{a}_{\mu} \lambda^{a}/2$. $g_s$ is the $SU_c(3)$ 
gauge coupling. $\lambda_a$ are Gell-Mann matrices in the color 
space, a runs from $1 \cdots 8$. $q_f=(u,d)^T$ denotes the quarks 
coming in two flavors and three colors. g is the flavor blind 
Yukawa coupling that couples the two flavor of quarks with four
mesons; one scalar ($\sigma, J^{P}=0^{+}$) and three pseudoscalars 
($\vec\pi, J^{P}=0^{-}$).

The quarks have no intrinsic mass but become massive after 
spontaneous chiral symmetry breaking because of nonvanishing 
vacuum expectation value of the chiral condensate. The mesonic 
part of the Lagrangian has the following form
\bea  
\label{eq:Lagmes}
{\cal L}_{m} & = &\frac 1 2 (\partial_\mu \sigma)^2+ \frac{ 1}{2}
  (\partial_\mu \vec \pi)^2 - U(\sigma, \vec \pi ) 
\eea

The pure mesonic potential is given by the expression
\be
U(\sigma,\vec{\pi})=\frac{\lambda}{4}\left(\sigma^2+\vec{\pi}^2-v^2\right)^2-h\sigma,
\ee
Here $\lambda$ is quartic coupling of the mesonic fields,
v is the vacuum expectation value of 
scalar field when chiral symmetry is explicitly broken 
and $h$ =$f_{\pi} m_{\pi}^2$ .

\subsection{Polyakov loop potential and thermodynamic grand potential}
\label{subsec:Plgtp}

The effective potential ${\cal U} \left( \Phi, \Phi^*, T \right)$ 
is constructed such that it reproduces thermodynamics of pure glue 
theory on the lattice for temperatures upto about twice the 
deconfinement phase transition temperature. In this work,we are using the  
logarithmic form of Polyakov loop effective potential Ref.~\cite{Ratti:07}.
The results produced by this potential are known to be fitted well to 
lattice results.

\bea
\label{eq:logpot}
\frac{{\cal U_{\text{log}}}\left(\Phi,\Phi^*, T \right)}{T^4} &=& -\frac{a\left(T\right)}{2}\Phi^* \Phi +
b(T) \, \mbox{ln}[1-6\Phi^* \Phi \nn \\&&+4(\Phi^{*3}+ \Phi^3)-3(\Phi^* \Phi)^2]
\eea

where the temperature dependent coefficients are as follow

\begin{equation*} 
  a(T) =  a_0 + a_1 \left(\frac{T_0}{T}\right) + a_2 \left(\frac{T_0}{T}\right)^2 \; \; \; 
  b(T) = b_3 \left(\frac{T_0}{T}\right)^3\ .
\end{equation*}

The critical temperature for deconfinement phase transition 
$T_0=270$ MeV is fixed for pure gauge Yang Mills theory.
In the presence of dynamical quarks $T_0$ is directly linked to the
mass-scale $\Lambda_{\rm QCD}$, the parameter which
has a flavor and chemical potential dependence in full dynamical QCD
and $T_0\to T_0(N_f,\mu)$ \cite{Schaefer:07,Herbst:2010rf}. For our 
numerical calculations in this paper, we have taken a fixed  $T_0=208$ for 
two flavours of quarks.

The parameters 
of Eq.(\ref{eq:logpot}) are 
\begin{eqnarray*}
&& a_0 = 3.51\ , \qquad a_1= -2.47\ , \nn \\ 
&& a_2 = 15.2\ ,  \qquad  b_3=-1.75\ 
\end{eqnarray*}

\begin{widetext}

In the mean-field approximation, the thermodynamic grand
potential for the PQM model is given as~\cite{Schaefer:07}

\begin{equation}
  \Omega_{\rm MF}(T,\mu;\sigma,\Phi,\Phi^*)  = {\cal
    U}(T;\Phi,\Phi^*) + U(\sigma ) +
  \Omega_{q\bar{q}} (T,\mu;\sigma,\Phi,\Phi^*). 
\label{Omega_MF}
\end{equation}

Here, we have written the vacuum expectation values $\langle \sigma \rangle = \sigma $ and $\langle \vec\pi \rangle =0$

The quark/antiquark  contribution in the presence of Polyakov loop reads

\begin{equation}
\Omega_{q\bar{q}} (T,\mu;\sigma, \Phi,\Phi^*) = \Omega_{q\bar{q}}^{\rm vac}+\Omega_{q\bar{q}}^{\rm T}
=- 2 N_f  \int
\frac{d^3 p}{(2\pi)^3} \left\{
 {N_c E_q} \theta( \Lambda^2 - \vec{p}^{\,2})  + T \Bigl[ \ln g_{q}^{+} + \ln g_{q}^{-} \Bigr]
\right\}
\label{Omega_MF_q}
\end{equation}

The first term of the Eq.~(\ref{Omega_MF_q}) denotes the fermion vacuum
contribution, regularized by the ultraviolet cutoff $\Lambda$.
In the second term $g_{q}^{+}$ and $g_{q}^{-}$ have been defined 
after taking trace over color space.

\bea
\label{eq:gpls} 
  g_{q}^{+} =  \Big[ 1 + 3\Phi e^{ -E_{q}^{+} /T} +3 \Phi^*e^{-2 E_{q}^{+}/T} +e^{-3 E_{q}^{+} /T}\Big] \,
\eea
\bea
\label{eq:gmns} 
  g_{q}^{-} =  \Big[ 1 + 3\Phi^* e^{ -E_{q}^{-} /T} +3 \Phi e^{-2 E_{q}^{-}/T} +e^{-3 E_{q}^{-} /T}\Big] \,
\eea

Here we use the notation E$_{q}^{\pm} =E_q \mp \mu $ and $E_q$ is the
single particle energy of quark/antiquark.
\be 
E_q = \sqrt{p^2 + m{_q}{^2}}
\ee 
where the constituent quark mass $m_q=g\sigma$
is a function of chiral condensate. In vacuum $\sigma(0,0) = \sigma_0 = f_\pi=  93.0 MeV$ 
\end{widetext}

\subsection{The renormalized vacuum term and model parameters}
\label{subsec:Vterm}
The fermion vacuum loop contribution can be obtained by appropriately renormalizing 
the first term of Eq.~(\ref{Omega_MF_q}) using the dimensional regularization scheme,
as done in Ref.\cite{Skokov:2010sf}. A brief description of essential steps is given below.

Fermion vacuum term is just the one-loop zero 
temperature effective potential at lowest order ~\cite{Quiros:1999jp}
\begin{eqnarray}
\label{vt_one_loop}
\Omega_{q\bar{q}}^{\rm vac} &=&   - 2 N_f N_c  \int \frac{d^3
  p}{(2\pi)^3} E_q  \nonumber\\ 
&=&    - 2 N_f N_c  \int \frac{d^4 p}{(2\pi)^4}  \ln(p_0^2+E_{q}^{2})
+ {\rm K}, 
\end{eqnarray}
the infinite constant $K$ is independent of
the fermion mass, hence it is dropped.

The dimensional regularization of Eq.~(\ref{vt_one_loop}) near three 
dimensions, $d=3-2\epsilon$ leads to the potential up to zeroth order in $\epsilon$ 
as given by
\begin{equation}
\Omega_{q\bar{q}}^{\rm vac} =  \frac{N_c N_f}{16 \pi^2} m_q^4 \left\{
  \frac{1}{\epsilon} - \frac{1}{2} 
\left[  -3 + 2 \gamma_E + 4 \ln\left(\frac{m_q}{2\sqrt{\pi} M}\right)
\right] 
\right\},
\label{Omega_DR}
\end{equation}
here $M$ denotes the arbitrary renormalization scale.

The addition of a counter term  $\delta \mathcal{L}$ in the Lagrangian of the QM or PQM model 

\begin{equation}
\delta \mathcal{L} = \frac{N_c N_f}{16 \pi^2} g^4\sigma^{4} \left\{ \frac{1}{\epsilon} - \frac{1}{2}
\left[  -3 + 2 \gamma_E - 4 \ln\left(2\sqrt{\pi}\right)  \right] \right\},
\label{counter}
\end{equation}

gives the renormalized fermion vacuum loop contribution as 
\begin{equation}
\Omega_{q\bar{q}}^{\rm reg} =  -\frac{N_c N_f}{8 \pi^2} m_q^4  \ln\left(\frac{m_q}{ M}\right).
\label{Omega_reg}
\end{equation}

Now the first term of Eq.~(\ref{Omega_MF_q}) which is vacuum contribution 
will be  replaced by the appropriately renormalized fermion vacuum loop contribution
as given in Eq.~(\ref{Omega_reg}).

The relevant part of the effective potential in Eq.~(\ref{Omega_MF}) which will fix 
the value of the parameters $\lambda$ and $v$ in the vacuum at $T=0$ and $\mu=0$ is the
purely $\sigma$ dependent mesonic potential $U(\sigma)$ plus the 
renormalized vacuum term given by Eq.~(\ref{Omega_reg}).
\begin{equation}
\Omega(\sigma) = \Omega_{q\bar{q}}^{\rm reg}+U(\sigma)
 = -\frac{N_c N_f}{8 \pi^2} g^4\sigma^4 \ln\left(\frac{g\sigma}{ M}\right)
- \frac{\lambda v^2}{2}\sigma^2+\frac{\lambda}{4}\sigma^4-h\sigma, 
\label{OmegaZEROT}
\end{equation}
The first derivative of $\Omega(\sigma)$ with respect to $\sigma$ at $\sigma=f_\pi$ in the vacuum is put to zero
\begin{equation}
\frac{\partial \Omega_{\rm MF}(0,0;\sigma,\Phi,\Phi^*)}{\partial\sigma}
= \frac{\partial \Omega(\sigma)}{\partial \sigma} =0
\label{DERIZERO}
\end{equation}
The second derivative of $\Omega(\sigma)$ with respect to $\sigma$ at $\sigma=f_\pi$ in the vacuum gives the mass of $\sigma$
\begin{equation}
m_\sigma ^2 = \frac{\partial^2 \Omega_{\rm MF}(0,0; f_\pi,\Phi,\Phi^*)}{\partial \sigma^2}
	    = \frac{\partial^2 \Omega(\sigma)}{\partial \sigma^2}
\label{msigma2}
\end{equation}

Solving the equations (\ref{DERIZERO}) and (\ref{msigma2}), we obtain 
\begin{equation}
\lambda = \lambda_s+\frac{N_c N_f}{8 \pi^2} g^4\left[3+4\ln\left(\frac{g f_\pi}{ M}\right)\right]
\label{lambd}
\end{equation}
and 
\begin{equation}
\lambda v^2 = (\lambda v^2)_s+\frac{N_c N_f}{4 \pi^2} g^4\ f_\pi^2 
\label{lambdv2}
\end{equation}
where $\lambda_s$  and  $(\lambda v^2)_s$ are the values of the parameters in the pure sigma model 
\begin{equation}
\lambda_s =\frac{m_\sigma^2-m_\pi^2}{2 f_\pi^2}
\label{lambs}
\end{equation}
\begin{equation}
(\lambda v^2)_s  =\frac{m_\sigma^2-3m_\pi^2}{2}
\label{lambv*2s}
\end{equation}

It is evident from the equations (\ref{lambd}) and (\ref{lambdv2}) that the value of
the parameters $\lambda$ and $v^2$ have a logarithmic dependence on the arbitrary 
renormalization scale M. However, when we put the value of $\lambda$ and $\lambda v^2$
in Eq.(\ref{OmegaZEROT}), the M dependence cancels out neatly after the rearrangement of terms.
Finally we obtain 

\begin{equation}
\Omega(\sigma) =  -\frac{N_c N_f}{8 \pi^2} g^4\sigma^4 \ln\left(\frac{\sigma}{f_\pi}\right)
- \frac{\lambda_r v_r^2}{2}\sigma^2+\frac{\lambda_r}{4}\sigma^4-h\sigma, 
\label{OmegasigF}
\end{equation}

Here, we define $\lambda_r$  and  $\lambda_r v_r^2$ as the values of the parameters after proper 
accounting of the renormalized fermion vacuum contribution.

\begin{equation}
\lambda_r = \lambda_s+\frac{3 N_c N_f}{8 \pi^2} g^4
\label{lambdR}
\end{equation}
and 
\begin{equation}
\lambda_r v_r^2 = (\lambda v^2)_s+\frac{N_c N_f}{4 \pi^2} g^4\ f_\pi^2 
\label{Rlambdv2}
\end{equation}

Now the thermodynamic grand potential for the PQM model in the presence of appropriately 
renormalized fermionic vacuum contribution (PQMVT model) will be written as 

\begin{equation}
\Omega_{\rm MF}(T,\mu;\sigma,\Phi,\Phi^*)  = {\cal
U}(T;\Phi,\Phi^*) + \Omega(\sigma ) + 
\Omega_{q\bar{q}}^{\rm T}(T,\mu;\sigma,\Phi,\Phi^*).
\label{OmegaMFPQMVT}
\end{equation}

Thus in the PQMVT model, One can get the chiral condensate $\sigma$, and the Polyakov
loop expectation values $\Phi$, $\Phi^*$ by searching the global 
minima of the grand potential in Eq.(\ref{OmegaMFPQMVT}) for a given value of temperature T 
and chemical potential 

\begin{equation}
  \frac{\partial \Omega_{\rm MF}}{\partial
      \sigma} =
  \frac{\partial \Omega_{\rm MF}}{\partial \Phi} = \frac{\partial
    \Omega_{\rm MF}}{\partial \Phi^*} =0\ ,
\label{EoMMF}
\end {equation}

We will take the values $m_\pi=138$ MeV, $m_\sigma=500$
MeV, and  $f_\pi=93$ MeV in our numerical computation. The constituent quark mass in 
vacuum $m_{q}^{0}=335$ MeV fixes the value of Yukawa coupling $g=3.3$.


\section{Effect of The Vacuum Term on The Phase Structure and Thermodynamics }
\label{sec:psvt}


%
We are presenting the results of our calculation for studying the temperature variation 
of the order parameters $\sigma$, $\Phi$, $\Phi^*$, their temperature derivatives and various 
thermodynamic observables at zero and non zero quark chemical potentials in the presence 
of the renormalized fermionic vacuum term in the effective potential of the PQM model. 
These results have been termed as PQMVT model calculations and we have investigated the 
interplay of chiral symmetry restoration and confinement-deconfinement transition in the 
influence of fermionic vacuum term. The phase diagram together with
the location of critical end point (CEP) has been obtained in $\mu$, and T plane for both the 
cases with and without fermionic vacuum contribution in the effective potential. 
In order to have a comparison, we have also shown the temperature variations of order 
parameters and their derivatives in the PQM model calculation with the same parameter set.
The temperature variations of thermodynamic observables namely pressure, energy density and
entropy density at three different chemical potentials (zero, $\mu_{CEP}$ and $\mu >\mu_{CEP}$) 
have been shown in PQMVT model calculations. In order to study the effect of fermionic vacuum 
term at zero chemical potential, the temperature variation of the interaction measure, speed of sound,
p/$\epsilon$ ratio and specific heat, has been calculated in PQMVT model and QMVT model 
(Quark Meson model with vacuum term) and these results have been compared  with the corresponding 
results in the PQM and QM model calculations. Finally we will be presenting the results of the temperature 
variation of baryon number density and quark number susceptibility at different chemical 
potentials in PQMVT model calculation.
\subsection{ Phase structure}
\label{subsec:Phastruct}
%
%

\begin{figure*}[!thbp]
\begin{center}
\begin{tabular}{ccc}
\includegraphics[width=0.5\textwidth]{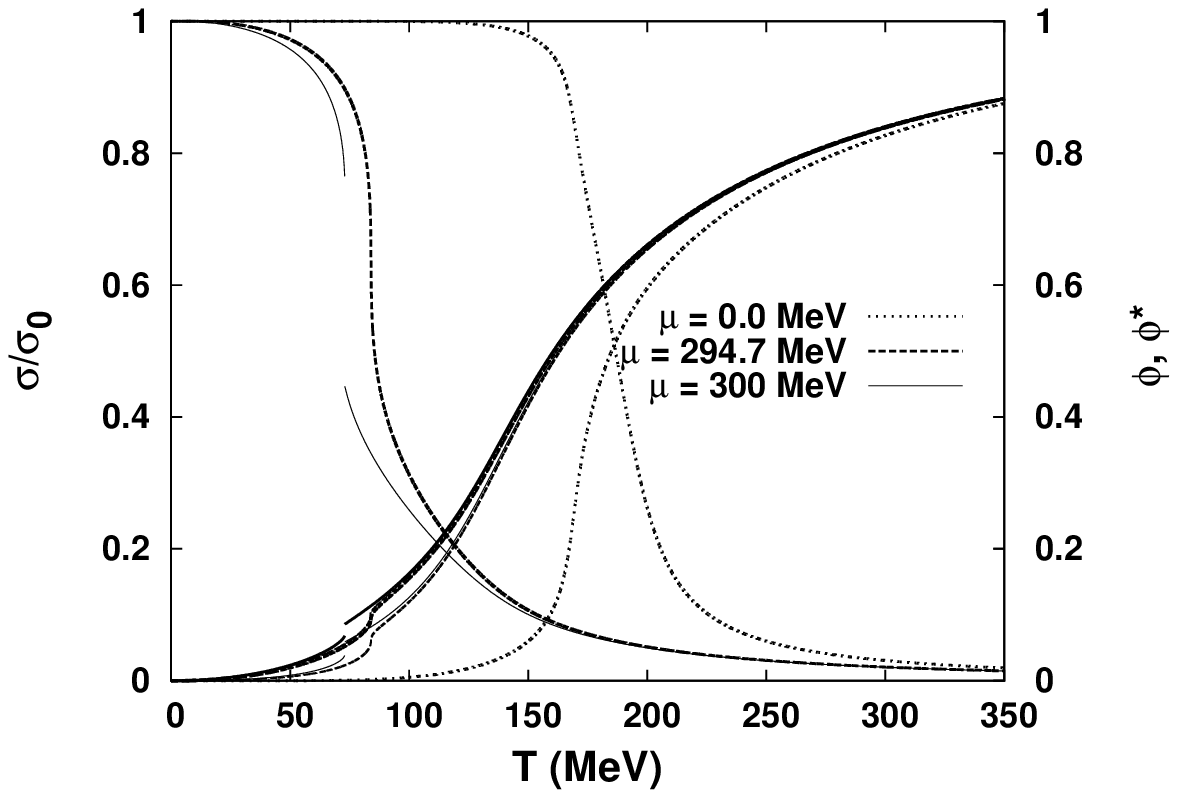} 
&&
 \includegraphics[width=0.45\linewidth]{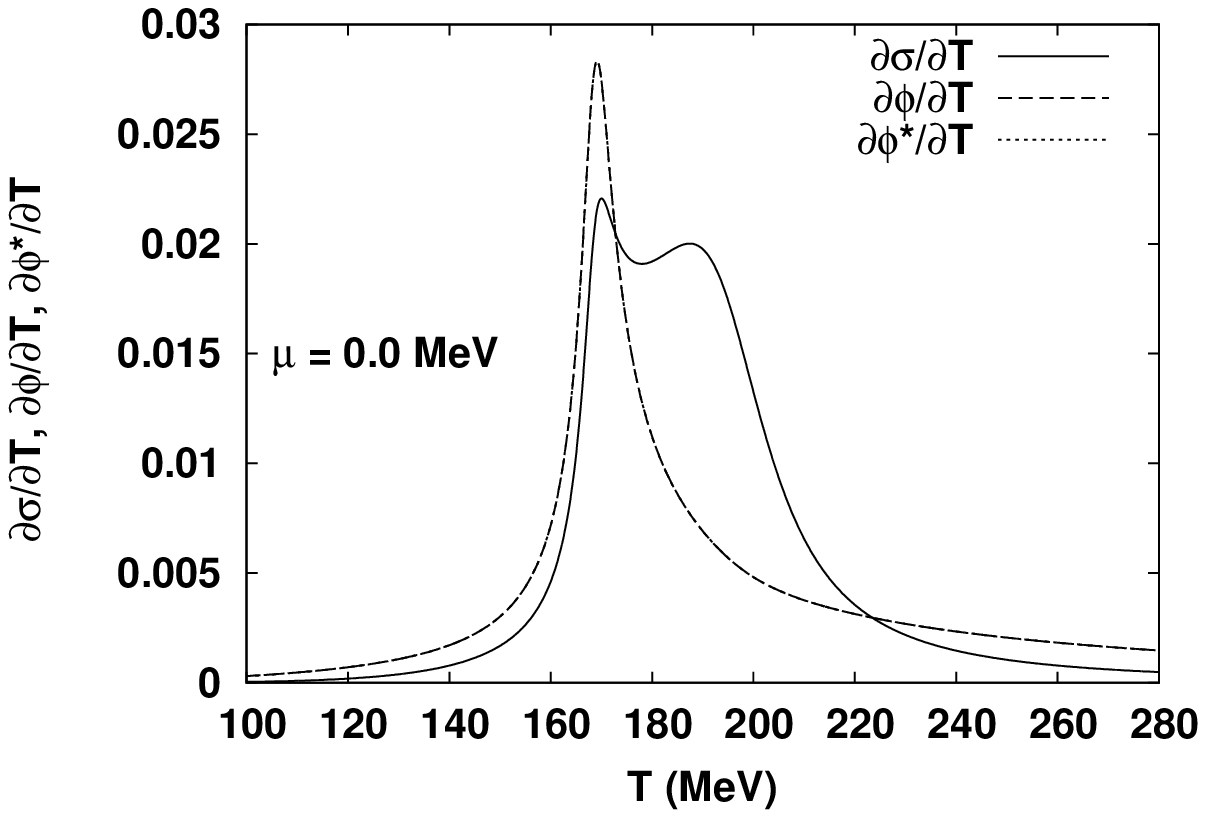} \\
(a) && (b)
\end{tabular}
\begin{tabular}{ccc}
\includegraphics[width=0.45\linewidth]{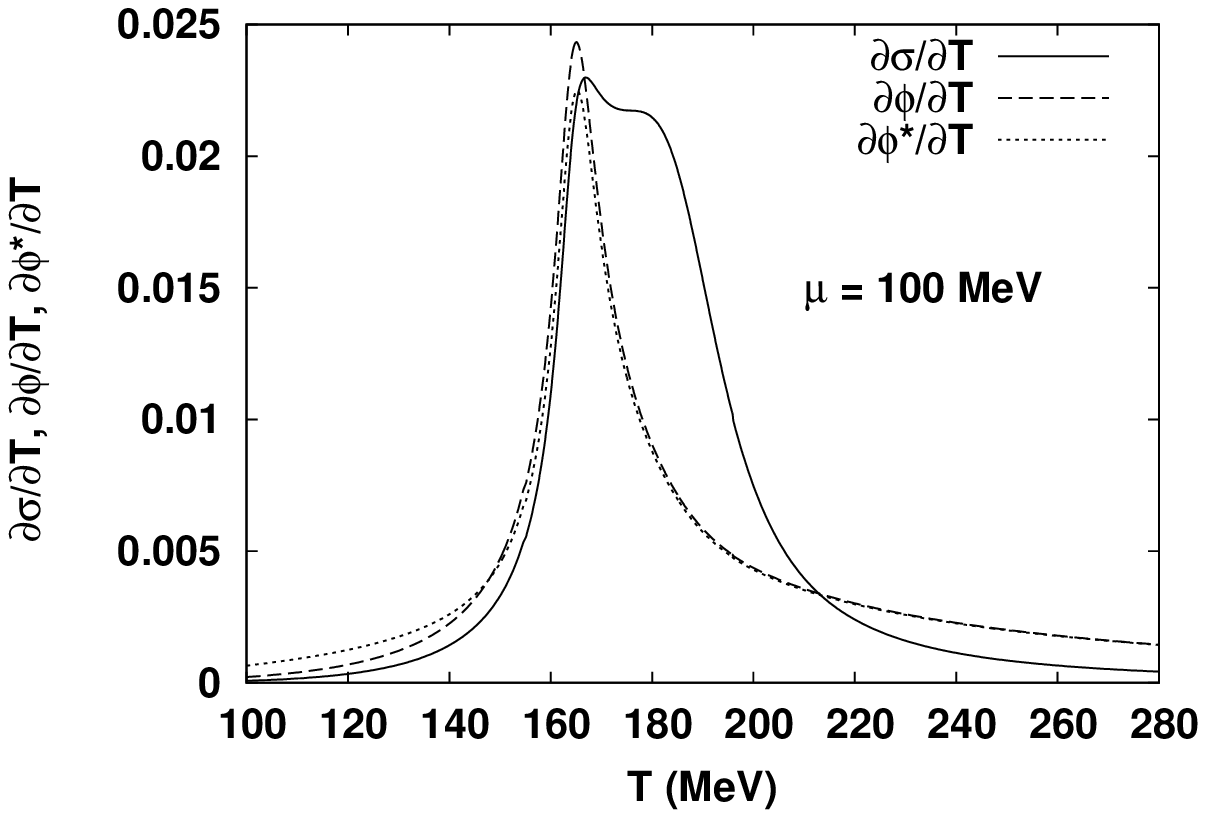}
&& 
\includegraphics[width=0.45\linewidth]{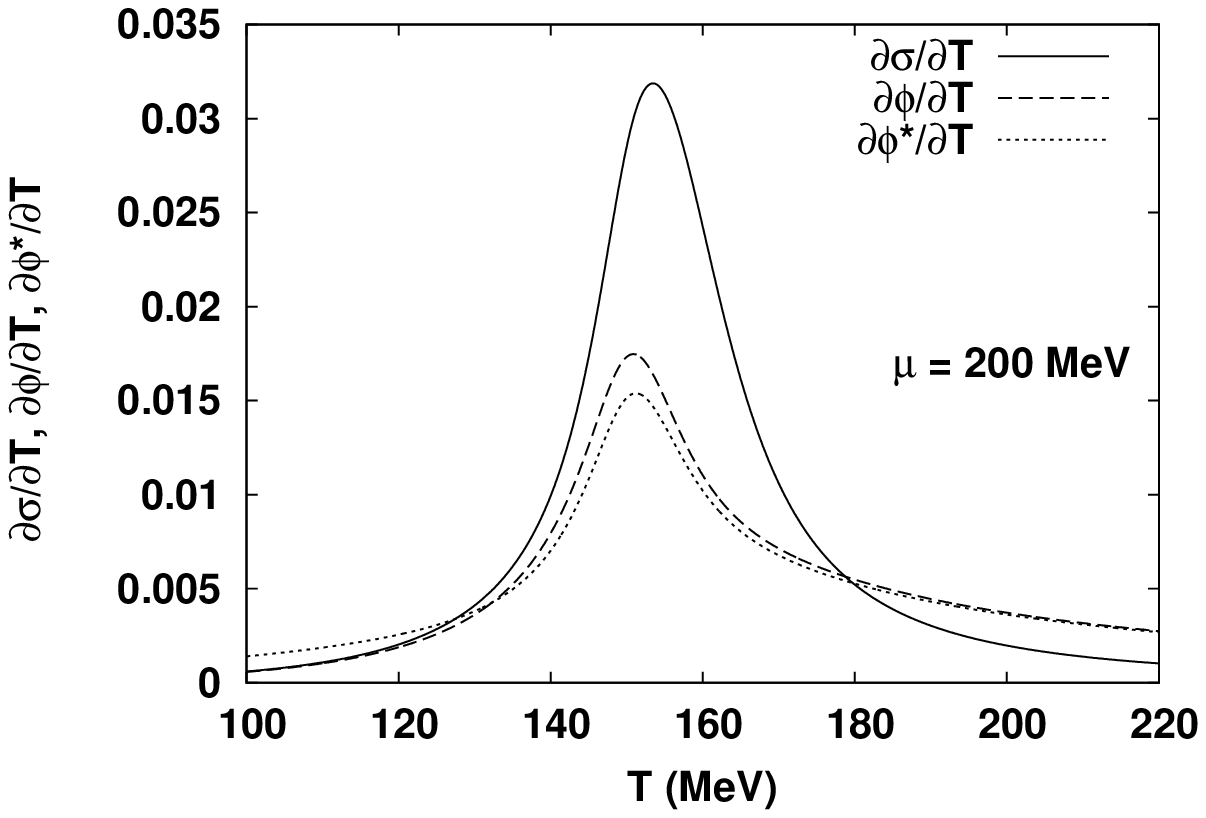} \\
(c) && (d)
\end{tabular}
\caption{Temperature variations in the PMQVT model. (a) The continuous dots , dash and solid lines represent 
the variation of $\frac{\sigma}{\sigma_0}$ on the left end and $\Phi$ on the right end of the plot at 
$\mu = 0$, $294.73$ (CEP) and $300$ MeV respectively. Thick dash and thick solid lines in the right end of 
the plot represent the $\Phi^*$ variations at $\mu = 294.7$ and $300$ MeV respectively.
(b), (c) and (d), show the temperature derivatives of $\sigma$, $\Phi$ and $\Phi^*$ fields
as a function of temperature respectively at three different chemical potentials 
$\mu$= 0, 100 and 200 MeV.}
\label{fig:fig1} 
\end{center}
\end{figure*}

The solutions of the coupled gap equations, Eq.(\ref{EoMMF}) 
determine the nature of chiral and deconfinement phase transition 
through the temperature and chemical potential dependence of 
chiral condensate $\sigma$,the expectation value of the Polyakov loop $\Phi$ 
and $\Phi^*$. Fig.1(a) shows the temperature variation of the 
chiral condensate $\sigma$ normalized with the vacuum value on the left
while the right end of the plot shows the Polyakov loop $\Phi$ and $\Phi^*$ temperature variation for the 
PQMVT model calculations, the corresponding temperature variation of the chiral and Polyakov loop order
parameters in PQM model calculations has been shown in Fig.2(a).
In Fig.1(a), the continuous dots, thin dash and thin solid lines represent the variation of $\frac{\sigma}{\sigma_0}$ on the 
left and $\Phi$ on the right at $\mu = 0$, $294.7$ (CEP) and $300$ MeV respectively. 
Thick dash and thick solid lines in the right end of the plot represent the $\Phi^*$ variations 
at $\mu = 294.7$ and $300$ MeV respectively. Fig.1(b), 1(c) and 1(d), show the temperature derivatives of $\sigma$,
$\Phi$ and $\Phi^*$ fields as a function of temperature respectively at three different chemical potentials 
$\mu = 0$ ,$100$ and $294.73$ MeV in PQMVT model calculations while the temperature variations of the same derivatives
in the PQM model at $\mu= 0$ MeV has been shown in Fig.2(b). The characteristic 
temperatures (pseudocritical temperature) for the chiral transition $T_{c}^{\chi}$ and the confinement-deconfinement 
transition $T_{c}^{\Phi}$,  are defined by the peak positions (inflection point) in the temperature derivatives 
of $\sigma$ and $\Phi$, $\Phi^*$ fields.

The chiral crossover transition for the realistic case of  explicitly broken chiral symmetry, becomes quite soft and 
smooth at  $\mu=0 $ because the corresponding chiral phase transition for massless quarks turns second order
in the chiral limit after having a proper accounting of the fermionic vacuum contribution in the PQMVT model.
The smoothness of crossover at $\mu=0$ is evident from the temperature variation of 
the chiral order parameter in Fig.1(a) while the Polyakov loop order parameter variation at the same chemical potential,
is sharp in comparison. The chiral crossover at  $\mu=0 $ becomes less smooth as we increase the chemical potential.
We find a large range ($\mu=0$ at $T_{c}^{\chi}=186.5$ MeV to $\mu_{CEP}=294.7$ MeV at $T_{c}^{\chi}=84.0$ MeV) in the values 
of chemical potential that makes the temperature variation of chiral order parameter, sharp and sharper such that 
eventually the crossover turns into a second order transition at CEP. 
The narrow width of the coincident variation of $\Phi$ and $\Phi^*$ temperature derivative at zero chemical potential
in Fig.1(b), signifies a sharp crossover for the confinement-deconfinement transition at $T_{c}^{\Phi}=169.0$ MeV 
The $\sigma$ derivative shows a broad double peak structure at $\mu=0$ similar to the findings of NJL model calculation 
in Ref\cite{Fu:07}, we have identified the chiral 
crossover temperature $T_{c}^{\chi}= 186.5$ MeV as the second peak position at higher pseudocritical temperature 
in Fig.1(b). The first peak in the $\sigma$ derivative is driven by the sharp peak of the Polyakov loop variation.
As the chemical potential is increased, the variation of Polyakov loop $\Phi$ derivative becomes smoother and broader with 
increasing width, while the $\sigma$ derivative variation shows a decreasing width and double peak
structure starts getting smeared after $\mu=100$ MeV as shown in Fig.1(c). For the chiral crossover transition 
in the chemical potential range $\mu= 100$ to $160$ MeV, the identification of pseudocritical 
temperature $T_{c}^{\chi}$ becomes ambiguous with an ambiguity of about 5 MeV. For $\mu >160$ MeV
in the PQMVT model, double peak structure disappears from the temperature variation of 
the chiral order parameter temperature derivative as shown in Fig.1(d) and its width decreases becoming narrow,
narrower and narrowmost till the CEP at $\mu = 294.7$ MeV and $T=84.0$ MeV is reached where the chiral 
transition turns second order.

For the realistic case of explicitly broken symmetry, the temperature variation of chiral order parameter at $\mu=0$, 
turns out to be quite sharp and rapid in Fig.2(a) in comparison to the corresponding PQMVT 
model variation in Fig.1(a) because the chiral transition in the massless quark limit, is first order in the PQM model. 
Further the chiral transition remains a crossover in quite a small range from $\mu=0$ at $T_{c}^{\chi}= 171.5$ MeV to 
$\mu=81$ MeV at $T_{c}^{\chi}=167$ MeV in the PQM model results of Fig.2(a). Since the chiral crossover is sharper 
than the confinement-deconfinement crossover in the PQM model calculations, the single peak of the $\sigma$ field temperature 
derivative in Fig.2(b) at $\mu= 0$ is narrower and a lot higher than the peak in the variation of 
temperature derivatives of $\Phi$ and $\Phi^*$. We have scaled the variation of $\Phi$ and $\Phi^*$ temperature 
derivatives in Fig.2(b) by a multiple of 5 which shows a very small double peak kind of structure.
We consider the chiral and confinement-deconfinement crossovers nearly coincident at 
$\mu=0$, $T_{c}^{\chi}= 171.5$ and we get exact coincidence as we move towards the
CEP ($T= 167.0$ MeV  and $\mu= 81.0$ MeV) of the model
where on account of the transition turning second order, we get highest and narrowmost peak.

In order to probe the issue of double peak structures emerging in Fig.1(b), 1(c), the temperature 
derivatives of  $\sigma$, $\Phi$ and $\Phi^*$ fields have been evaluated as function of temperature 
by taking the Polynomial form \cite{Ratti:06} for Polyakov loop potential instead of the logarithmic 
form in PQMVT model. AT $\mu=0$, none of the field derivatives have double peak structure. It starts
appearing at $\mu =200$ MeV separately in $\Phi$ and $\Phi^*$ derivatives as shown in Fig.3(a) and we 
find robust noncoincident second peaks respectively at $T_{c}^{\Phi}= 176$ MeV, $T_{c}^{\Phi^*}= 156$ MeV 
for $\mu =280$ MeV as shown in Fig.3(b). The first and highest peak in $\sigma$ field temperature  derivative 
gives the location of CEP at $T_{CEP}=78.2 $ MeV and $\mu_{CEP}= 293$ MeV. Though we have not evaluated the whole phase diagram 
for this case, we find that the confinement-deconfinement transition line (obtained from nearly coincident peaks in 
$\Phi$ and $\Phi^*$) lies below the chiral crossover transition line ($T_{c}^{\Phi} < T_{c}^{\chi}$) 
in the chemical potential range $\mu = 0$ to $\mu = 200$ MeV and confinement-deconfinement crossover 
transitions for $\Phi^*$ and $\Phi$ fields for $\mu > 200$ MeV separate in constituting different
lines which get located above the chiral crossover phase boundary
from $\mu =200$ to $\mu = \mu_{CEP}= 293$ MeV . These findings being similar to the results of 
Ref.\cite{kahara,Herbst:2010rf} 
support the quarkyonic phase\cite{larry} like scenario, having a region of confinement with restored 
chiral symmetry. It has been argued in Ref.\cite{kahara} that transition temperature in the chiral sector decreases 
as the chemical potential is increased while the remnant of the deconfinement transition 
remain uneffected by the value of chemical potential. We notice that this explanation does not work 
in a calucation  with logarithmic ansatz for Polyakov loop potential. Thus the meaning of the peaks in $\Phi$, 
$\Phi^*$ and $\sigma$ field temperature derivatives and their relation to the nature of crossover transition in 
confinement-deconfinement and chiral sector is debatable. Similar to the findings of NJL model calculation 
in Ref\cite{Fu:07}, our calculations in PQMVT model with phenomenologically improved logarithmic Polyakov loop potential 
(which describes the gluon dynamics more appropriately with a better correlation to the effect of dynamical quarks),
show double peak in $\sigma$ field temperature derivative while the Polyakov field $\Phi$ derivative shows a 
single peak in a chemical potential range  $\mu=0$ to $\mu=200$ MeV as discussed in the second paragraph.
Here we find support for the standard scenario \cite{Herbst:2010rf,Schaefer:07} where chiral symmetry restoration occurs 
at a higher temperature than the deconfinement transition. Thus the quarkyonic phase scenario is ruled out in our PQMVT 
model calculation with constant $T_0$ when we have taken a logarithmic form for the Polyakov loop potential while 
the calculation with polynomial Polyakov loop potential supports its occurrence in certain range of $\mu$ and $T$.

\begin{figure*}[!tbp]
\subfigure[]{
\label{fig:mini:fig2:a} 
\begin{minipage}[]{0.5\linewidth}
\centering \includegraphics[width=\linewidth]{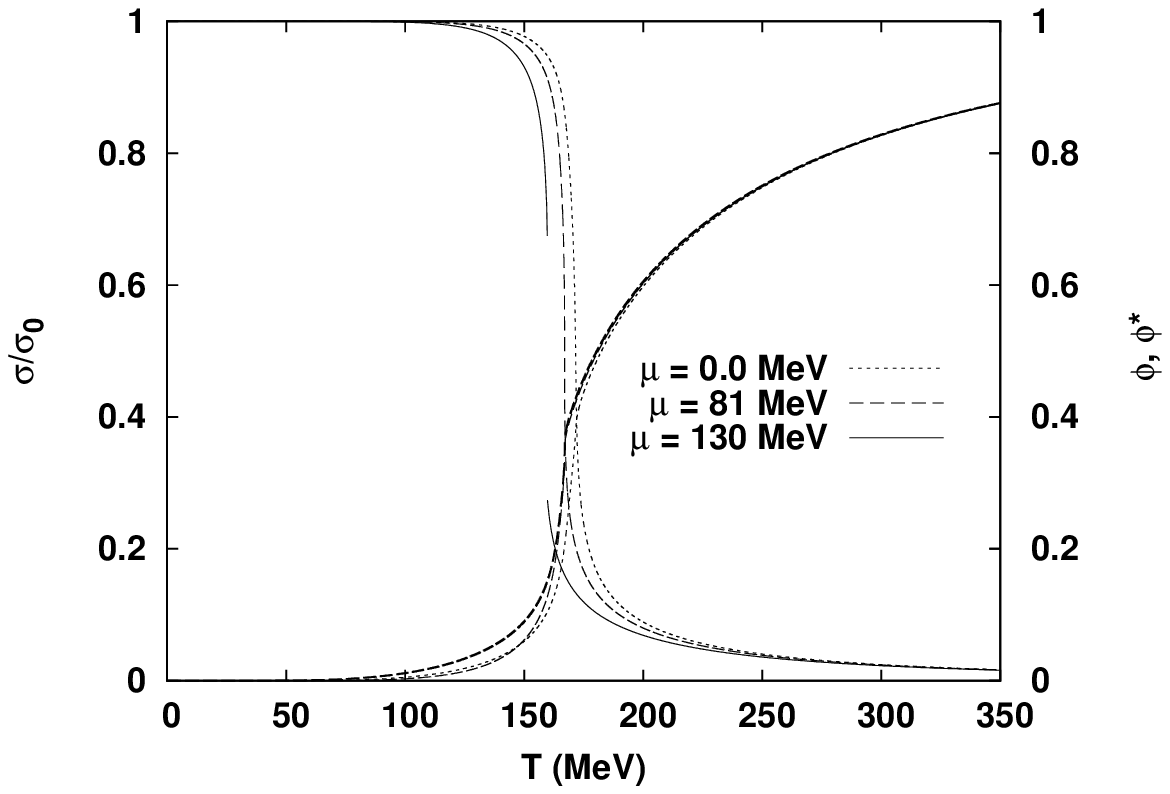}
\end{minipage}}%
\hspace{0.25in}
\subfigure[]{
\label{fig:mini:fig2:b} 
\begin{minipage}[]{0.4\linewidth}
\centering \includegraphics[width=\linewidth]{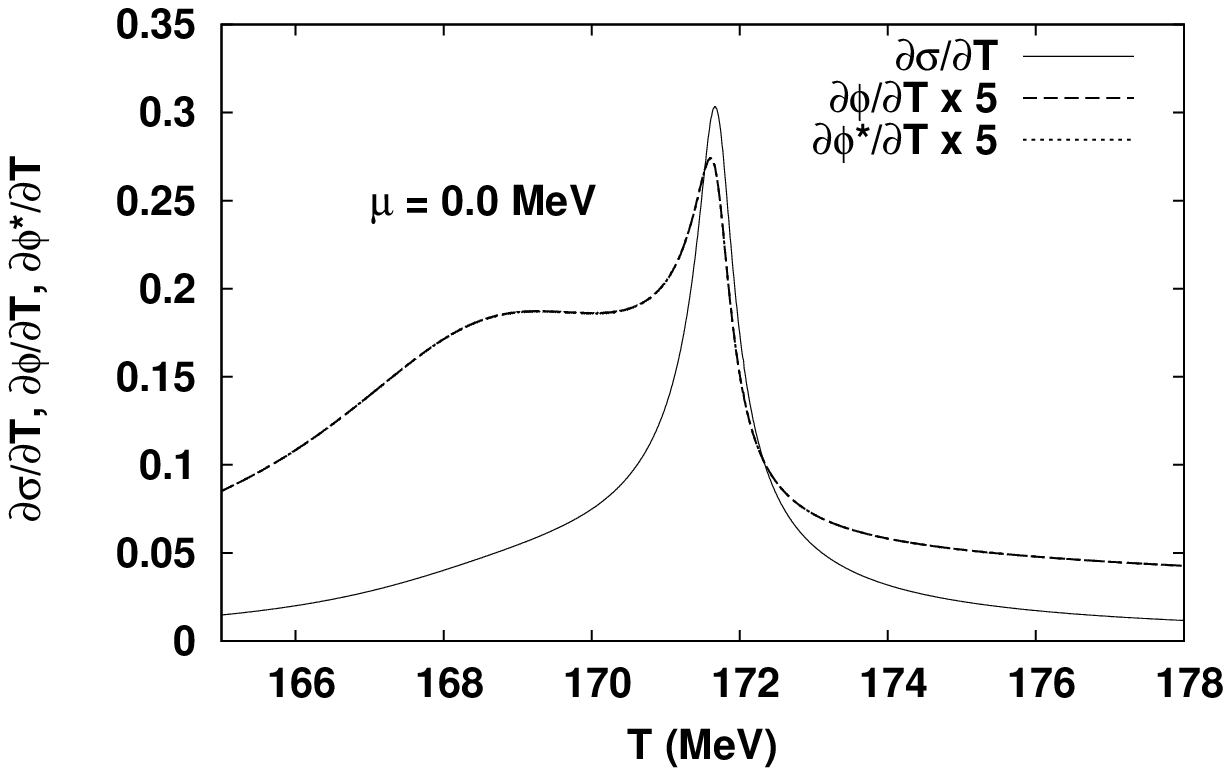}
\end{minipage}}
\caption{(a) The continuous dots, dash and solid lines in the left half of the figure represent the variation of 
$\frac{\sigma}{\sigma_0}$ in the PQM model at $\mu = 0$, $\mu = 81$ and $\mu = 130$ MeV respectively. In the 
right end of the plot, 
continuous dots represent coincident variation of $\Phi$ and $\Phi^*$ at $\mu = 0$ while thick and thin dash 
lines represent the $\Phi^*$ and $\Phi$ variations at $\mu = 81$  MeV respectively.
(b) shows the temperature derivatives of $\sigma$, $\Phi$ and $\Phi^*$ fields
as a function of temperature at $\mu= 0$ in the PQM model.}
\label{fig:mini:fig2} 
\end{figure*}

\begin{figure*}[!tbp]
\subfigure[]{
\label{fig:mini:fig3:a} 
\begin{minipage}[]{0.4\linewidth}
\centering \includegraphics[width=\linewidth]{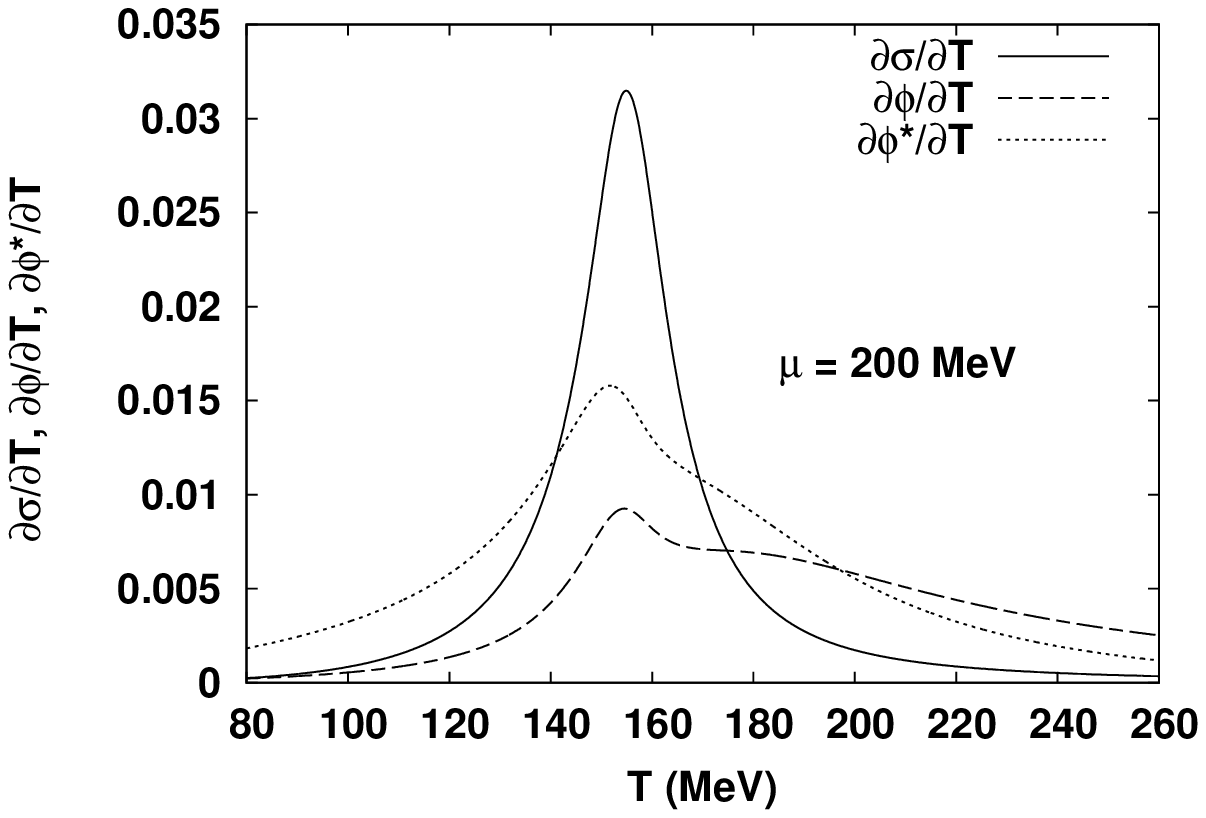}
\end{minipage}}%
\hspace{0.24in}
\subfigure[]{
\label{fig:mini:fig3:b} 
\begin{minipage}[]{0.4\linewidth}
\centering \includegraphics[width=\linewidth]{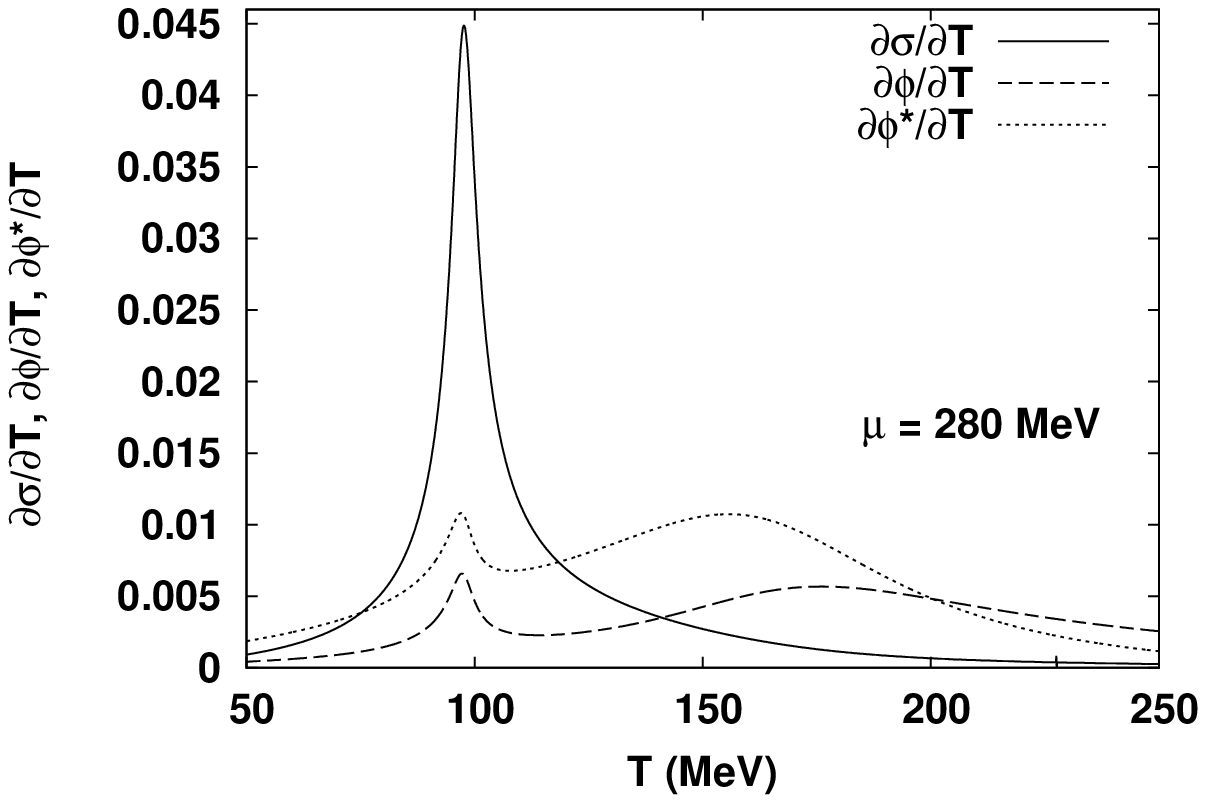}
\end{minipage}}
\caption{Temperature variation of order parameter derivatives with polynomial Polyakov loop 
potential in PQMVT model calculation.}
\label{fig:mini:fig3} 
\end{figure*}

\begin{figure*}[!hbt]
\centering
\hspace{-0.3cm}\includegraphics[width=0.6\linewidth]{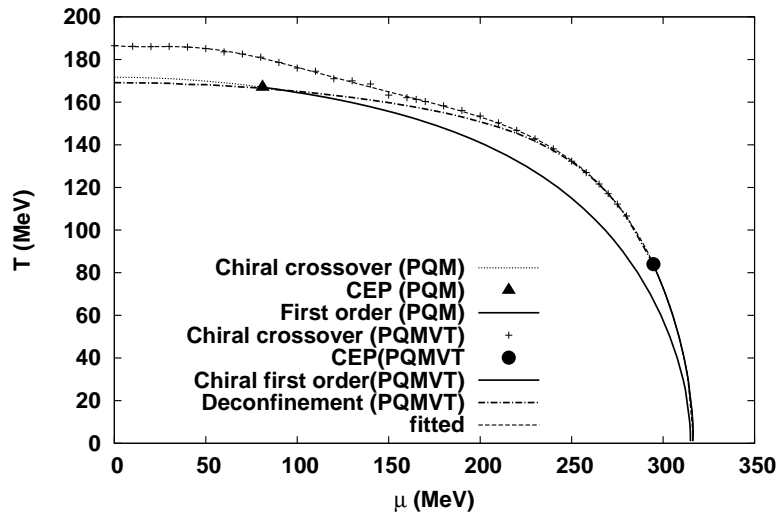}
\caption{Phase Diagram.}
\label{fig:fig4}
\end{figure*}

In Fig.4, we have obtained the phase diagram in our calculation with logarithmic Polyakov loop potential and 
located the critical end point (CEP) in the PQMVT as well as PQM model calculations 
for $m_{\sigma}= 500$ MeV. The structure of the phase diagram is very sensitive to the chosen value of the sigma meson mass. 
For the value $m_{\sigma}= 600$ MeV, the transition becomes a crossover in the entire $\mu$ and T plane for the PQMVT model 
calculation. We have shown the chiral crossover transition by a dash line  which starts from $T_{c}^{\chi} =186.5$ MeV 
at $\mu=0$ axis and ends at CEP; $T_{CEP} =84$ MeV and $\mu_{CEP}=294.7$ MeV in PQMVT model. Due to the smearing 
of  double peak structure in the temperature derivative of chiral order parameter in the range $\mu = 100$ to $160$ MeV,
the chiral crossover transition temperature $T_{c}^{\chi}$ is identified with an ambiguity of about 5 MeV.
We get a unique $T_{c}^{\chi}$ for $\mu >160$ MeV in the phase diagram because of a single peak structure
which gets narrow and narrower for higher chemical potentials till we reach the CEP. The dash dotted line 
which starts at $T_{c}^{\Phi}=169$ MeV and ends at CEP of the PQMVT model, signifies the confinement-
deconfinement crossover transition. The chiral  and confinement-deconfinement crossover transition 
lines merge at $\mu= 250 $ and $T_{c}^{\chi}  = T_{c}^{\Phi}= 132$. The thin solid line
for $\mu > \mu_{CEP}$ represents the first order phase transition corresponding to the jump in all
the order parameters at the same critical temperature. The chiral crossover transition line lies 
above the crossover line for the confinement-deconfinement transition. Thus our results of the PQMVT
model calculation  are in tune with the standard scenario \cite{Herbst:2010rf} where chiral symmetry restoration 
occurs at a higher critical temperature $T_{c}^{\chi} =186.5$ MeV than the confinement-
deconfinement transition temperature $T_{c}^{\Phi}=169 $ MeV at $\mu=0$ axis. Further  the 
crossover transition temperature at $\mu=0$ compare well with the lattice \cite{Aoki:06,Cheng:06} results
and QCD based computations \cite{Herbst:2010rf,Braun:2009gm} in two flavour model.
The chiral  and confinement-deconfinement crossover transition lines are coincident (as shown by the continuous dots)  and 
start from $T_{c}^{\chi}=T_{c}^{\Phi}=171.5$ MeV at $\mu=0 $ MeV to end at the CEP of the PQM model. The first order transition  
for $\mu >\mu_{CEP}$ in the PQM model, has been shown by the thick solid line. The CEP of the PQM model gets located near 
the temperature axis at $\mu_{CEP}=81$ MeV and $T_{CEP}=167$ MeV because the chiral crossover at $\mu=0 $, having the background of a 
first order phase transition in the chiral limit, is rapid and sharp and soon it gets converted to a first order phase transition 
as we increase the chemical potential. While the critical end point (CEP) gets shifted close to the chemical potential axis in 
PQMVT model because the chiral crossover transition at $\mu = 0$ is quite soft and smooth as it emerges from a phase transition 
which turns second order in the chiral limit due to the effect of renormalized fermionic vacuum contribution in the effective 
potential and further it remains a crossover for large values of the chemical potential.

\subsection{Thermodynamic Observables:Pressure,Entropy and Energy Density}
\label{subsec:Topeed}

\begin{figure*}[!ht]
\centering
\hspace{-0.3cm}\includegraphics[width=0.6\linewidth]{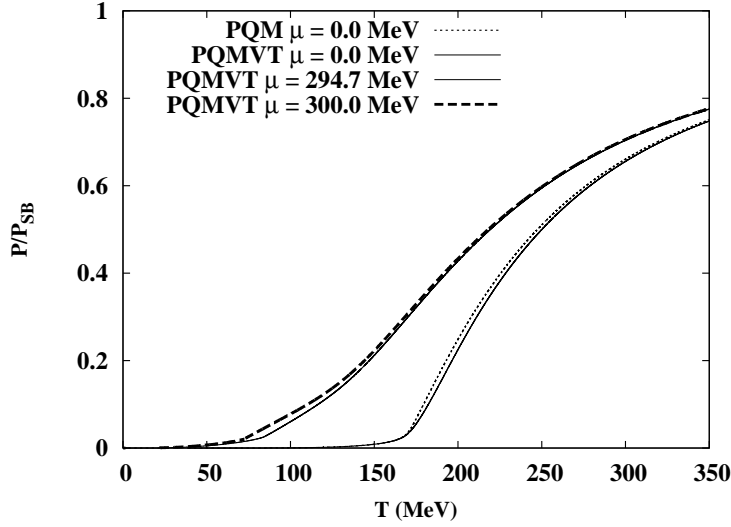}
\caption{Pressure variation with respect to temperature.}
\label{fig:fig5}
\end{figure*}

The negative of grand potential gives the thermodynamic pressure 

\begin{equation}
p(T,\mu) = - \Omega_{\rm MF}\left(T,\mu \right)
\end{equation}
Thermodynamic pressure divided by the QCD Stefan-Boltzmann (SB) limit has been shown for three chemical potentials 
$\mu= 0, 294.7$ (CEP) and 300 MeV in Fig.5 for PQMVT model. It has been normalized to vanish at $T=\mu=0 $. We have shown the 
the pressure calculated in PQM model also for comparison at $\mu =0 $. For $N_{f}$ massless quarks and $N_{c}^2-1$ massless 
gluons in the deconfined phase, the QCD pressure in the SB limit is given by 

\begin{equation}
  \frac{ p_{\mathrm{SB}}}{T^4} = (N_c^2-1)\frac{ \pi^2}{45}
  + N_c N_f\!\!
  \left[ \frac{ 7\pi^2}{180} \!+\! \frac{ 1}{6} \left( 
      \frac{ \mu}{T}
    \right)^2 \!+\! \frac{ 1}{12\pi^2} \left( \frac{ \mu}{T} 
    \right)^4
  \right]\ .  
\end{equation}

The fermionic vacuum contribution makes the pressure variation in PQMVT model smooth at $\mu= 0$ and 
this curve (thin solid line) lies slightly below the curve (line with continuous dots) obtained in PQM model. 
The  pressure variations at  $\mu_{CEP} =294.7$ and $\mu= 300$ MeV of PQMVT model are represented by the thick solid and dash 
line respectively.The pressure increases near the chiral transition due to the melting of the constituent quark masses 
and saturates at about eighty percent of the SB limit. 

The entropy density is defined as negative of the temperature derivative of the grand potential. 

\begin{equation}
 s = - \frac{\partial\Omega_{\rm MF}}{\partial T} 
\label{eq:entropdens}
\end{equation}
The implicit variation of $\sigma$, $\Phi$ and $\Phi^*$ fields with
respect to temperature has been accounted for, in the temperature derivative of $\Omega(\sigma)$,
$\cal U_{\text{log}}$ and $\Omega_{\mathrm{q\bar{q}}}^{\rm T}$
as evaluated in the appendix A.
The temperature variation of entropy density normalized by its QCD SB limit has been shown in Fig.6. It is 
continuous for crossover transition and attains about 40-45 percent of its SB value at psedocritical transition temperature.  
Again due to the fermionic vacuum fluctuations, the entropy density variation (thin solid line) at $\mu=0$ turns out 
to be a smoother function of temperature in PQMVT model when it is compared with corresponding curve (line with
continuous dots) of PQM model calculation.  At $\mu_{\textrm{CEP}}= 294.7$ MeV, the entropy density curve (thick solid line) 
shows a steep rise at $T_{\textrm{CEP}}=84$. MeV in PQMVT model, then it takes a bend to reach its saturation. 
The PQM model entropy density curve (dah dotted line) at $\mu= 294.7$ MeV shows a large jump because chiral 
transition is strong first order at this chemical potential. The first order chiral transition of PQMVT model at 
$\mu= 300.0$ MeV, generates another  jump in the entropy density curve (line with dash), though this jump is smaller 
than the  first order jump seen in PQM  model entropy curve at a lower chemical potential $\mu= 294.7$ MeV.

\begin{figure*}[!ht]
\centering
\hspace{-0.3cm}\includegraphics[width=0.4\linewidth,angle=-90]{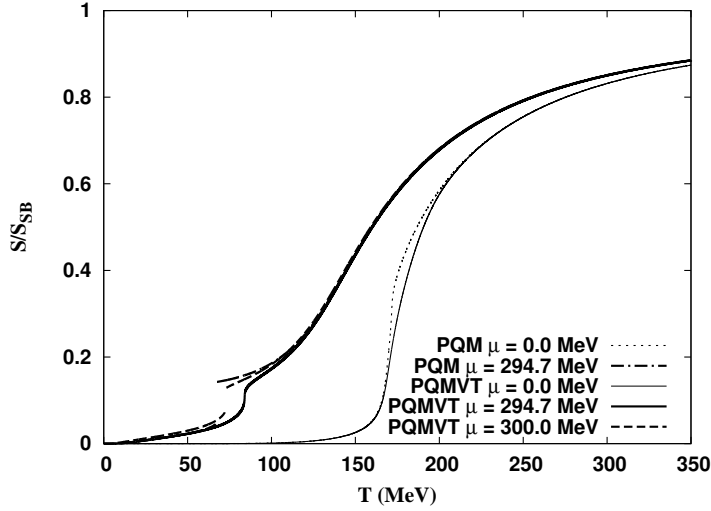}
\caption{Entropy variation with respect to temperature.}
\label{fig:fig6}
\end{figure*}

\begin{figure*}[!ht]
\centering
\hspace{-0.3cm}\includegraphics[width=0.4\linewidth,angle=-90]{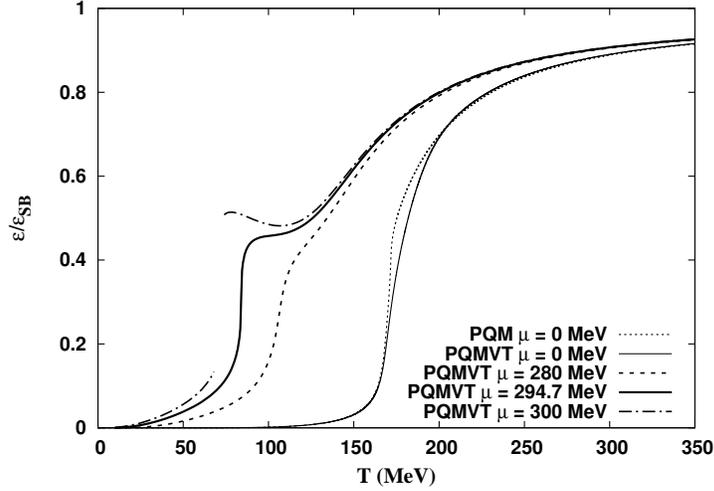}
\caption{Variation of energy density with respect to temperature.}
\label{fig:fig7}
\end{figure*}

The energy density in the presence of chemical potential is given as 
\begin{equation}
\label{eq:energydensity}
\epsilon = -p + T s +\mu n 
\end{equation}
where n is the number density.The  temperature variation of energy density normalized by its QCD SB limit value
has been shown in Fig.7 for $\mu= 0 , 280$ , $294.7$ (CEP) and $300$ MeV in PQMVT model. The energy density 
variation (thin solid line) at $\mu= 0$ similar to entropy density variation, is smoother in comparison 
to the corresponding variation in PQM model calculation (line with continuous dots), this again is 
due to the influence of fermionic vacuum fluctuations. Similar to the entropy density variation
at $\mu_{\textrm{CEP}}= 294.7$ MeV, the energy density (thick solid line) also shows a very steep and large 
rise at $T_{\textrm{CEP}}=84.0$ MeV, then it curves to attain the saturation. At $\mu= 300.0$ MeV, we get a large jump in 
the energy density curve (dash-dotted line) which of course is a signature of the first order chiral transition. 
Since quark degrees of freedom get liberated and become light, the energy density registers a rapid increase near 
the crossover/phase transition point and reaches almost to the value of SB limit.

\begin{figure*}[!ht]
\centering
\hspace{-0.3cm}\includegraphics[width=0.4\linewidth,angle=-90]{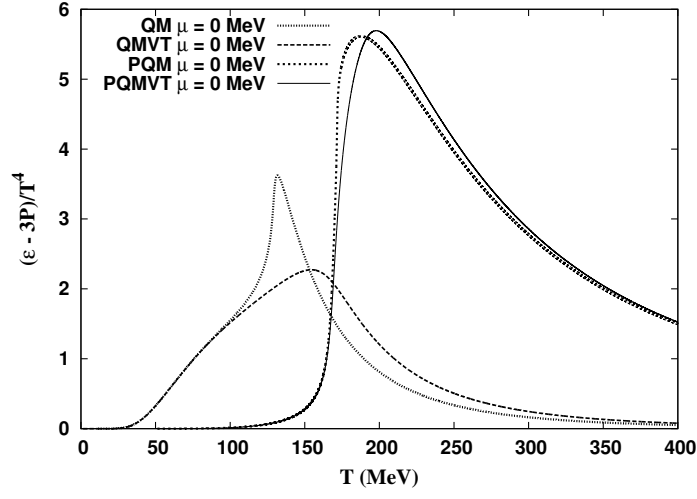}
\caption{Change in interaction measure with respect to temperature.}
\label{fig:fig8}
\end{figure*}

The trace anomaly of energy momentum tensor is also known as interaction measure.
The temperature variation of the interaction measure $\triangle =(E-3p)$/$T^4$ has been shown in Fig.8 at 
$\mu=0$ MeV in PQMVT, QMVT and PQM, QM model calculations. The QM model variation of the interaction measure (line with 
continuous dots) shows a sharp and narrow peak near the pseduocritical transition temperature which becomes very 
broad and smooth in the corresponding variation ( thick dash line)  of QMVT model calculation due to the effect 
of inclusion of fermion vacuum term contribution in the effective potential of QM model. The peak of interaction 
measure temperature variation (solid line) in PQMVT model shifts to a slightly higher temperature value in comparison 
to the corresponding peak in the variation (line with thin dash) of PQM model calculation.  

\subsection{Specific heat $C_{V}$ and Speed of sound $C_{S}$} 
\label{subsec:sheat}
\begin{figure*}[!ht]
\centering
\hspace{-0.3cm}\includegraphics[width=0.55\linewidth]{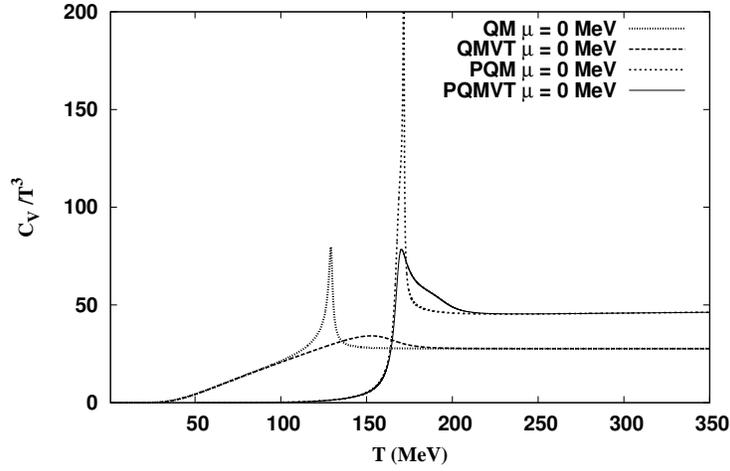}
\caption{Specific heat variation with respect to temperature.}
\label{fig:fig9}
\end{figure*}

 The expression of specific heat at constant volume is given by 
\begin{equation} 
\label{eq:cv}
C_V = \left.\frac{\partial \epsilon}{\partial T}\right|_V = -T \left.
  \frac{\partial^2 \Omega_{\rm MF}}{\partial T^2} \right|_V 
\end{equation}

The second partial temperature derivatives of $\sigma$, $\Phi$ and $\Phi^*$ fields
contribute in the double derivatives of $\Omega(\sigma)$,
$\cal U_{\text{log}}$ and $\Omega_{\mathrm{q\bar{q}}}^{\rm T}$
with respect to temperature as given in the appendix A.
Fig.9 shows the temperature variation of the specific heat $C_{V}$ normalized by $T^3$ in QM, QMVT and in 
PQM, PQMVT model calculations at $\mu=0$. The specific heat variation while growing with the temperature, peaks at the 
crossover transition temperature and then saturates at the corresponding SB limit at the higher temperature.
The QM model specific heat variation shows a large and sharp peak which becomes quite smooth and broad 
in the corresponding variation of QMVT model calculation due to the presence of fermionic vacuum term, 
further the peak position gets shifted to a higher transition temperature. The qualitative difference of structures 
in the curves of QM and QMVT model gets reduced due to the influence of Polyakov loop potential and we notice that the 
PQM model specific heat variation has quite a high and sharp peak which becomes small and a little 
less sharp in the PQMVT model variation and the 
peaks occur at the same transition temperature. Further, we remark that the peak positions of the temperature 
variation of order parameter derivatives in Fig.1(b) and Fig.2(b) give different transition temperatures for chiral crossover in PQM and PQMVT model 
calculations while for confinement-deconfinement crossover, the transition temperature is almost same in both the models.
 
The speed of sound is an important quantity for hydrodynamical investigations of relativistic heavy-ion collisions.
It is given by 
\begin{equation}
  C_s^2 = \left.\frac{\partial  p}{\partial \epsilon}\right|_S =
  \left.\frac{\partial p}{\partial T}\right|_V \left/
    \left.\frac{\partial \epsilon}{\partial T}\right|_V = \frac{
      s}{C_V} \right., 
\end{equation}
The equation of state parameter $p(T)/\epsilon(T)$ also represents the information contained in trace anomaly. 
The velocity of sound $C_{s}^2 $  and the equation of state parameter $p(T)/\epsilon(T)$ ratio has been shown 
as a function of temperature in QM, QMVT and PQM, PQMVT model calculations at $\mu = 0$ in Fig.10. Thick lines 
denote the result for the sound velocity $C_{s}^2 $ and thin lines show the variation of $p(T)/\epsilon(T)$ ratio. 
The presence of fermion vacuum term in QMVT model leads to a very smooth temperature variation for $C_{s}^2$ (line with thick long
dash) and $p(T)/\epsilon(T)$ ratio (line with thin long dash). The $C_{s}^2$ temperature variation (line with thick 
continuous dots) in the QM model calculation, shows a very sharp drop followed by a rapid rise while the EOS parameter 
$p(T)/\epsilon(T)$ ratio (line with thin continuous dots) shows a cusp  at crossover transition temperature. The 
temperature variation of $C_{s}^2$ (thick solid line) and $p(T)/\epsilon(T)$ ratio (thin solid line) in the PQMVT 
model turns out to be smoother than the corresponding variation of $C_{s}^2$ (line with thick,short and dark dash) 
and $p(T)/\epsilon(T)$ ratio (thin dash line) in the PQM model calculation. At higher temperatures $C_{s}^2 $ and 
$p(T)/\epsilon(T)$ ratio approach the ideal gas value $1/3$ in all the cases of model calculation. 
In PQM and PQMVT models, the value of $C_{s}^2$ almost matches with the $p(T)/\epsilon(T)$ ratio for lower 
temperatures and $C_{s}^2$ value is always larger than the $p(T)/\epsilon(T)$ ratio accept near the transition 
temperature, similar as in Ref.\cite{Schaefer:09ax,Tamal:06}. The minimum value of $p(T)/\epsilon(T)$ ratio is 
about .033 in PQMVT model which being slightly larger than the PQM model value .026, is less than the lattice 
result .075 \cite{Ejiri:2005uv,Cheng:08}.  Similar to the findings of Ref.\cite{Schaefer:09ax}, interestingly 
the $C_s^2$ value is found to be  less than 
0.1 around half the crossover transition temperature in our PQM and PQMVT model results. In contrast,using a 
model of confinement, the results of  Ref.~\cite{Mohanty:2003va} find values of about $C_s^2 \sim 0.2$ around 
half the transition temperature and $C_s^2 =0.15$ around the transition temperature.

\begin{figure*}[!ht]
\centering
\hspace{-0.3cm}\includegraphics[width=0.5\linewidth]{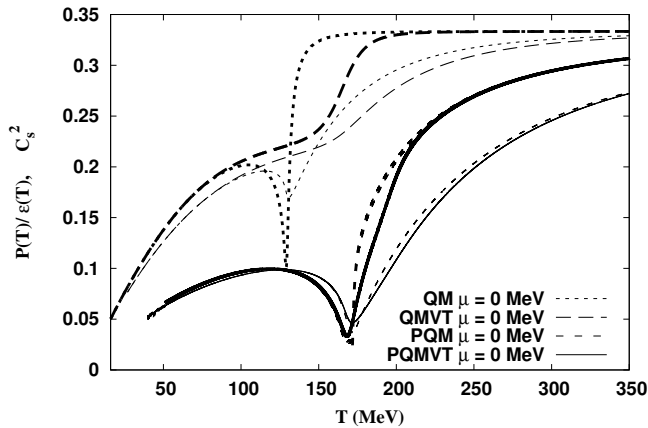}
\caption{The variation of $p(T)/\epsilon(T)$ has been shown by thin lines while thick lines show 
the variation of $C_{s}^{2}$.}
\label{fig:fig10}
\end{figure*}

%
%
\subsection{ Quark number density and Susceptibility }
\label{subsec:Qnqs}
 
\begin{figure*}[!ht]
\centering
\hspace{-0.3cm}\includegraphics[width=0.5\linewidth]{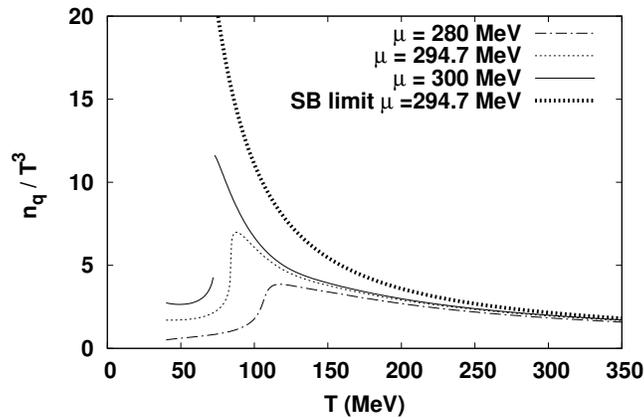}
\caption{Temperature variation of quark number density divided by $T^3$.}
\label{fig:fig11}
\end{figure*}

The first derivative of grand potential with respect to chemical potential gives the quark number density
 \begin{equation}
 n = - \frac{\partial\Omega_{\rm MF}}{\partial \mu} 
\label{eq:qndens}
\end{equation}

The implicit variation of $\sigma$, $\Phi$ and $\Phi^*$ fields with
respect to chemical potential has been accounted for, in the evaluation of first derivative of 
$\Omega(\sigma)$, $\cal U_{\text{log}}$ and $\Omega_{\mathrm{q\bar{q}}}^{\rm T}$
with respect to chemical potential as given in the appendix A.
The temperature variation of the quark number density normalized by $T^3$ in the PQMVT model 
calculation has been shown in Fig.11 for three quark chemical potentials $\mu = 280 , 294.7$ 
(CEP) and $300$ MeV. The dash dotted line shows the number density variation 
for a crossover transition at $\mu = 280$ MeV, here we see a small peak structure. The dotted line  number density variation shows a sharper rise with a narrow peak  at $\mu_{\textrm{CEP}} = 294.7$ MeV and it approaches the SB value of number density variation (shown by thick dots) for higher temperatures. At $\mu = 300$ MeV, the Phase transition becomes first order, hence the quark number density
being a first derivative of the grand potential with respect to the chemical potential, shows a jump in the solid line temperature variation.

The expression of quark number susceptibility  is obtained as 
\begin{equation}
 \chi_{q}=\frac{\partial^2 \Omega_{\rm MF}}{\partial \mu^2} 
\label{eq:qsuscept}
\end{equation}

The second partial derivatives of $\sigma$, $\Phi$ and $\Phi^*$ fields with respect to chemical
potential contribute in the double derivatives of $\Omega(\sigma)$,
$\cal U_{\text{log}}$ and $\Omega_{\mathrm{q\bar{q}}}^{\rm T}$
with respect to chemical potential as given in the appendix A.
Fig.12 shows the variation of quark number susceptibility normalized by $T^2$ as a function of temperature in the PQMVT model calculation for chemical potentials $\mu = 280 , 294.7$ (CEP) and $300$ MeV. The dash dotted line susceptibility variation
 at $\mu = 280$ MeV  shows a continuous peak structure at the  crossover transition temperature.
Since at $\mu_{\textrm{CEP}} = 294.7$ MeV, the phase transition turns second order, the dotted line of quark 
number susceptibility variation shows a very large and strongly divergent peak at $T_{\textrm{CEP}}$. The solid line shows 
the quark number susceptibility at $\mu = 300$ MeV for the first order transition case, we get a discontinuous variation because order parameter registers a jump in the first order transition.     

\begin{figure*}[!hbt]
\centering
\hspace{-0.3cm}\includegraphics[width=0.5\linewidth]{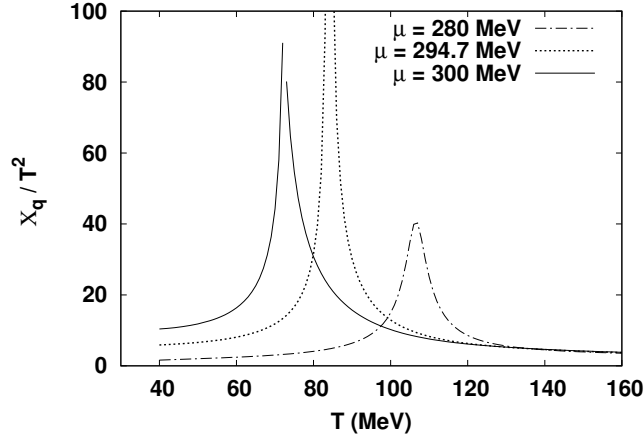}
\caption{Susceptibility $\chi_q /T^2$ variation with respect to temperature.}
\label{fig:fig12}
\end{figure*}
%
%
\section{Summary and discussion}
\label{sec:smry}

We have investigated the temperature variation of the order parameters $\sigma$, $\Phi$, $\Phi^*$, their temperature 
derivatives and various thermodynamic physical observables at non zero and zero quark chemical potentials in
the presence of renormalized fermionic vacuum term in the effective potential of the PQM model.
The results termed as the PQMVT model calculations have been compared  with the results of PQM model 
without vacuum term. We have used logarithmic Polyakov loop potential for our calculation.

The chiral crossover transition for the realistic case of explicit chiral symmetry breaking, becomes quite soft and 
smooth at  $\mu=0$ in PQMVT model due to the proper accounting of the fermionic vacuum term contribution in the  
PQM model because the corresponding phase transition at $\mu=0$ turns second order in the chiral limit of 
massless quarks. The $\sigma$ derivative shows a broad double peak structure at $\mu=0$. The second peak position 
at higher transition temperature $T_{c}^{\chi}= 186.5$ MeV identifies the chiral crossover because the first peak results
due to a sharp peak in the Polyakov loop temperature variation which signals a rapid confinement-deconfinement
crossover transition at $\mu=0$. In a large range of $\mu$, T values (from $\mu=0$ and $T=186.5$ MeV 
to $\mu=294$ MeV and $T=84$ MeV), the chiral transition remains a crossover and it keeps on becoming sharper with the increase 
in chemical potential till the point of second order transition at $\mu_{CEP}$ is reached in the PQMVT model. Instead of 
logarithmic form, if we take polynomial form for Polyakov loop potential in our PQMVT model calculation, 
the temperature derivatives of Polyakov loop field $\Phi$ and its conjugate $\Phi^*$ show distinct non coincident double peak structure in the chemical potential range $\mu=200$ MeV to $\mu_{CEP}=293$ MeV and we do not find any double peak structure 
near $\mu=0$ in the temperature derivative of $\sigma$ field. Hence confinement-deconfinement crossover transition lines for $\Phi^*$ and $\Phi$ fields will be located above the chiral crossover phase boundary from $\mu =200$ to $\mu = \mu_{CEP}=293$ MeV.
This finding support the quarkyonic phase like scenario, having a region of confinement with restored chiral symmetry.
Our calculation with logarithmic form for the Polyakov loop potential does not support this quarkyonic phase like scenario.
Since the chiral transition in the massless quark limit is first order at zero chemical potential, the corresponding 
crossover transition for the realistic case has been found to be quite sharp and rapid in the PQM model 
without any vacuum term. Further the chiral transition remains a crossover in quite a small range only 
from $\mu=0$ and $T_c^{\chi}=171.5$ MeV to $\mu=81$ MeV and $T_c^{\chi}=167$ MeV in the PQM model calculations.

The phase diagram together with the location of critical end point (CEP) has been obtained in $\mu$, and T 
plane for $m_{\sigma}= 500$ MeV in both the models PQMVT and PQM. The structure of the phase diagram is very 
sensitive to the value of sigma meson mass. For the value $m_{\sigma}$= 600 MeV, the transition becomes a 
crossover in the entire $\mu$ and T plane for the PQMVT model calculation. We do not have a coincident chiral and 
confinement-deconfinement crossover transitions in the PQMVT model as the chiral crossover transition line lies
above the crossover line for the confinement-deconfinement transition. Our results of the PQMVT model calculation, 
are in tune with the standard scenario where chiral symmetry restoration occurs at a higher critical temperature 
than the confinement-deconfinement transition temperature. The quarkyonic phase scenario is ruled out in our PQMVT model calculation 
with constant $T_0$ when we have taken a logarithmic form for the Polyakov loop potential while the 
calculation with polynomial form of Polyakov loop potential supports its occurrence in certain range 
of $\mu$ and T. The critical end point (CEP) gets shifted close to the chemical potential axis 
($\mu_{CEP}=294.7$ MeV, $T_{CEP}=84.0$ MeV ) in PQMVT model because the chiral crossover transition 
at $\mu = 0$ emerging from a second order phase transition in the chiral limit, becomes quite soft and 
smooth due to the effect of fermionic vacuum contribution in the effective potential and further it 
remains a crossover for large values of the chemical potential. The chiral  and confinement-deconfinement 
crossover transition lines are coincident and the CEP of the PQM model gets located near the temperature 
axis at $\mu_{CEP}= 81 $  and $T_{CEP}=167 $ because the chiral crossover at $\mu=0 $, having the 
background of a first order phase transition in the chiral limit, is quite rapid and sharp and soon it 
gets converted to a first order phase transition as we increase the chemical potential.

The temperature variation of thermodynamic observables namely pressure, energy density, entropy
density at three different chemical potentials (zero, $\mu_{CEP}$ and $\mu > \mu_{CEP}$)  has been
shown in PQMVT model. Due to the proper accounting of appropriately renormalized fermionic vacuum fluctuations, the pressure ,entropy density and energy density  variations at $\mu=0$ turn out to be a smoother function of temperature in PQMVT model 
when it is compared with corresponding curves in PQM model calculation. The temperature variations of the interaction measure, 
speed of sound, $p(T)/\epsilon(T)$ and specific heat, have been calculated in PQMVT model and QMVT (Quark Meson model
with vacuum term) model and these results have been compared with the corresponding results in the PQM and QM model
calculations. Again we find that the presence of fermionic vacuum contribution in effective potential leads to the smoother 
variation of the thermodynamic quantities. Finally we have shown the results of the temperature variations of baryon number
density and quark number susceptibility at different chemical potentials in PQMVT model calculations.

%
\begin{acknowledgments}
We are very much thankful to Krysztof Redlich for an immensely helpful and fruitful 
discussion during the visit to  ICPAQGP-2010 at Goa in India. Valuable suggestions 
together with compuational helps by Rajarshi Ray during the completion of this work 
are specially acknowledged. General physics discussions with Ajit Mohan Srivastava
were very helpful.  We are also thankful to Ananta Prasad Mishra, Saumia PS, Ranjita 
Mohapatra, Abhishek Atreya, Biswanath Layek and Neelima Agarwal for valuable suggestions. 
We acknowledge the financial support of the Department of Atomic Energy- Board of Research
in Nuclear Sciences (DAE-BRNS), India, under the research grant No. 2008/37/13/BRNS.
We also acknowledge the computational support of the computing facility
which has been developed by the Nuclear Particle Physics group of the Physics
Department, Allahabad University under the Center of Advanced Studies(CAS)
funding of UGC India. 
\end{acknowledgments}


\appendix
\section{First and second partial derivatives of grand potential}
First partial derivative of logarithmic Polyakov loop potential with respect to chemical potential and temperature



\bea
\frac{\del {\cal U_{\text{log}}}}{\del \mu} &=&  \mathrm{T}^4 \Big[ -\frac{\mathrm{a(T)}}{2} \Big\{
 \frac{\del \Phi }{\del \mu} \Phi^* + \Phi \frac{\del\Phi^*}{\del \mu} \Big\} 
 -6 \mathrm{b(T)}~ \mathrm{X_{\mu}} \Big]
\eea
\vspace{-0.5cm}
\bea
\frac{\del {\cal U_{\text{log}}}}{\del \mathrm{T}} &=& 4 \mathrm{T}^3 \Big[-\frac{\mathrm{a(T)}}{2}\Phi^* \Phi 
+ \mathrm{b(T)} \, \mbox{ln}[W] \Big] 
+ \mathrm{T}^4 \Big[ -\frac{1}{2} \Big\{ \frac{\rm{d} \mathrm{a(T)}}{\rm{d} \mathrm{T}} \Phi \Phi^* 
+ \mathrm{a(T)} \frac{\del \Phi }{\del \mathrm{T}} \Phi^* +\mathrm{a(T)} \Phi \frac{\del\Phi^*}{\del \mathrm{T}} \Big\} \nn \\
&& +\frac{\rm{d} \mathrm{b(T)}}{\rm {d} \mathrm{T}} \mbox{ln} \left[\mathrm{W} \right] -6 \mathrm{b(T)}~ \mathrm{X_{T}} \Big]
\eea
\vspace{-0.5cm}

where

\be
X_{y} = \frac{\left( 1 + \Phi \Phi^* \right) \left( \Phi \frac{\del \Phi^*}{\del \mathrm{y}} 
+ \Phi* \frac{\del \Phi}{\del \mathrm{y}} \right) - 2 \left( \Phi^2 \frac{\del \Phi}{\del \mathrm{y}} 
+ {\Phi^*}^2 \frac{\del \Phi^*}{\del \mathrm{y}} \right) }{\mathrm{W}} 
\ee

\vspace{-0.5cm}

\be
W = 1-6\Phi^* \Phi +4(\Phi^{*3}+ \Phi^3)-3(\Phi^* \Phi)^2
\ee
 
Second partial derivative of logarithmic Polyakov loop potential with respect to chemical potential and temperatures

\bea
\frac{{\del}^2{\cal U_{\text{log}}}}{\del \mu^2} &=& 
 \mathrm{T}^4 \bigg[-\frac{\mathrm{a(T)}}{2} \Big( \frac{{\del}^2\Phi }{\del \mu^2} \Phi^*
+2 \frac{{\del}\Phi }{\del \mu} \frac{{\del}\Phi^* }{\del \mu}+\Phi \frac{{\del}^2\Phi^*} {\del \mu^2} \Big) 
-36 \mathrm{b(T)}  \mathrm{X_{\mu}}^2  \nn \\
&& -  \frac{6 \mathrm{b(T)}}{W} \bigg\{ \left( \Phi \frac{\del\Phi^*}{\del\mu}  
+\Phi^* \frac{\del\Phi}{\del \mu} \right)^2
+ \big( 1 + \Phi \Phi^* \big) \left( \Phi \frac{{\del}^2\Phi^*}{\del \mu^2} 
+2 \frac{\del\Phi}{\del\mu} \frac{\del\Phi^*}{\del\mu}
+ \Phi^* \frac{{\del}^2\Phi}{\del \mu^2} \right) \nn \\
&& - 2 \left(\Phi^2 \frac{{\del}^2\Phi}{\del\mu^2} 
+2 \Phi \left( \frac{\del \Phi}{\del\mu} \right)^2
 +{\Phi^*}^2 \left(\frac{\del^2 \Phi^*}{\del \mu^2}\right) 
+2 \Phi^* \left( \frac{\del\Phi^*}{\del\mu} \right)^2  \right) 
 \bigg\} \bigg]
\eea

\bea
\frac{{\del}^2{\cal U_{\text{log}}}}{\del T^2} &=& 12 \mathrm{T}^2 \Bigl[-\frac{\mathrm{a}(\mathrm{T})}{2}\Phi^* \Phi 
+ \mathrm{b}(\mathrm{T}) \, \mbox{ln}[\mathrm{W}] \Bigr] \nn  \\
&& + 8 \mathrm{T}^3 \bigg[
 -\frac{1}{2} \Big( \frac{\rm{d} \mathrm{a}(\mathrm{T})}{\rm{d} \mathrm{T}} \Phi \Phi^*
+ \mathrm{a}(\mathrm{T}) \frac{\del \Phi}{\del \mathrm{T}}\Phi^* 
+ \mathrm{a}(\mathrm{T}) \Phi \frac{\del \Phi^*}{\del \mathrm{T}} \Big) 
+\frac{\rm{d} \mathrm{b}(\mathrm{T})}{\rm{d} \mathrm{T}} \mbox{ln} \left[\mathrm{W} \right] 
- 6 \mathrm{b(T)} X_T \bigg]  \nn  \\
&& + \mathrm{T}^4 \bigg[-\frac{1}{2} \Big( \frac{{\rm{d}}^2\mathrm{a(T)}}{\rm{d} \mathrm{T^2}} \Phi \Phi^*
+ \mathrm{a(T)} \frac{{\del}^2\Phi }{\del \mathrm{T}^2} \Phi^*+ \mathrm{a(T)} \Phi \frac{{\del}^2\Phi^*}
{\del \mathrm{T}^2} \Big) \nn \\
&& \qquad \quad - \Bigl( \frac{\rm{d} \mathrm{a}(\mathrm{T})}{\del \mathrm{T}} \frac{\del \Phi}{\del \mathrm{T}} \Phi^*
+ \frac {\rm{d}\mathrm{a}(\mathrm{T})}{\rm{d} \mathrm{T}} \Phi \frac{\del \Phi^*}{\del \mathrm{T}}  
+ \mathrm{a}(\mathrm{T}) \frac{\del \Phi}{\del \mathrm{T}} \frac{\del\Phi^*}{\del\mathrm{T}} \Bigr)
+ \frac{{\rm{d}}^2\mathrm{b(T)}}{\rm{d} \mathrm{T}^2} \mbox{ln} \left[\mathrm{W} \right] \bigg] \nn  \\
&& -12 \frac{\rm{d}\mathrm{b(T)}}{\rm{d}\mathrm{T}} \mathrm{X_T} 
-36 \mathrm{b(T)}~ \mathrm{X_T}^2 \nn   \\
&& - 6 \mathrm{b(T)} \bigg[ \bigg\{ \left( \Phi \frac{\del\Phi^*}{\del\mathrm{T}}  
+\Phi^* \frac{\del\Phi}{\del \mathrm{T}} \right)^2
+ \big( 1 + \Phi \Phi^* \big) \left( \Phi \frac{{\del}^2\Phi^*}{\del \mathrm{T}^2} 
+2 \frac{\del\Phi}{\del\mathrm{T}} \frac{\del\Phi^*}{\del\mathrm{T}}
+ \Phi^* \frac{{\del}^2\Phi}{\del \mathrm{T}^2} \right) \nn \\
&& \qquad \quad 
 - 2 \left(\Phi^2 \frac{{\del}^2\Phi}{\del\mathrm{T}^2} 
+2 \Phi \left( \frac{\del \Phi}{\del\mathrm{T}} \right)^2
 +{\Phi^*}^2 \left(\frac{\del^2 \Phi^*}{\del \mathrm{T}^2}\right) 
+2 \Phi^* \left( \frac{\del\Phi^*}{\del\mathrm{T}} \right)^2  \right) 
 \bigg\} \frac{1}{W} \bigg] \bigg]
\eea

First partial derivative of $\Omega_{q\bar{q}}^{\rm T}$ with respect to chemical potential and temperature




\bea
\frac{\del \Omega_{\mathrm{q\bar{q}}}^{\rm T}}{\del \mu} &=& -12 \int \frac{\mathrm{d}^3\mathrm{p}}{\left(\mathrm{2 \pi}\right)^3}
 \Big[ \mathrm{T} \left(B_{q, \mu}^{+} + B_{q, \mu}^{-} \right) \Big]
\eea

\bea
\frac{\del \Omega_{\mathrm{q\bar{q}}}^{\rm T}}{\del \mathrm{T}} &=& -4 \int \frac{\mathrm{d}^3\mathrm{p}}{\left(\mathrm{2 \pi}\right)^3}
 \Big[ \ln g_{q}^{+} + \ln g_{q}^{-} +  3\mathrm{T} \left(B_{q,T}^{+} + B_{q,T}^{-} \right) \Big]
\eea

where

\begin{equation}
A_{q}^{+} = \Phi \mathrm{e}^{-\beta\mathrm{E_{q}}^{+}} +2 \Phi^* \mathrm{e}^{-2\beta\mathrm{E_{q}}^{+}} 
+\mathrm{e}^{-3\beta\mathrm{E_{q}}^{+}}  
\end{equation}
\begin{equation}
A_{q}^{-} =\Phi^*\mathrm{e}^{-\beta \mathrm{E_{q}}^{-}} +2 \Phi \mathrm{e}^{-2\beta \mathrm{E_{q}}^{-}} 
+\mathrm{e}^{-3\beta \mathrm{E_{q}}^{-}} 
\end{equation}

\be
B_{q,\rm{x}}^{+} = \frac{1}{ g_{q}^{+}} \Big\{ A_{q}^{+} \frac{\del}{\del \rm{x}}(-\beta \mathrm{E_{q}}^{+} ) 
+\frac{\del \Phi}{\del\mathrm{x}} \mathrm{e}^{-\beta \mathrm{E_{q}}^{+}} 
+\frac{\del \Phi^*}{\del\mathrm{x}} \mathrm{e}^{-2\beta \mathrm{E_{q}}^{+}} \Big\}  
\ee
\be
B_{q,\rm{x}}^{-} = \frac{1}{ g_{q}^{-}} \Big\{ A_{q}^{-} \frac{\del}{\del \rm{x}}(-\beta \mathrm{E_{q}}^{-} ) 
+\frac{\del \Phi^*}{\del\mathrm{x}} \mathrm{e}^{-\beta \mathrm{E_{q}}^{-}} 
+\frac{\del \Phi}{\del\mathrm{x}} \mathrm{e}^{-2\beta \mathrm{E_{q}}^{-}} \Big\}  
\ee

Second partial derivative of $\Omega_{q\bar{q}}^{\rm T}$ with respect to chemical potential and temperature
\bea
\frac{\del^2 \Omega_{\mathrm{q\bar{q}}}^{\rm T}}{\del\mu^2} &=& -12 \mathrm{T}\int\frac{\mathrm{d}^3\mathrm{p}}{\left(\mathrm{2\pi}\right)^3}
   \mathrm{D}_{\rm{q},\mu} 
\eea

\bea
\frac{\del^2 \Omega_{\mathrm{q\bar{q}}}^{\rm T}}{\del \mathrm{T}^2} &=& -12 \int \frac{\mathrm{d}^3\mathrm{p}}{\left(\mathrm{2 \pi}\right)^3}
\Big[2 \left( B_{q}^{+} + B_{q}^{-} \right) + \mathrm{T}~ \mathrm{D_{q,T}} \Big]
\eea

where
\bea
\mathrm{D_{q,x}} &=& \bigg[ -3 \left( {B_{q}^{+}}^2 + {B_{q}^{-}}^2 \right) \nn \\
&& +\frac{1}{g_{q}^{+}} \bigg\{C_{q}^{+} \Big[\frac{\del}{\del \rm{x}}(-\beta \mathrm{E_{q}}^{+} )\Big]^2 
+\left(2 \frac{\del \Phi}{\del\mathrm{x}} \mathrm{e}^{-\beta \mathrm{E_{q}}^{+}} 
+ 4 \frac{\del \Phi^*}{\del\mathrm{x}} \mathrm{e}^{-2\beta \mathrm{E_{q}}^{+}}\right)\frac{\del}{\del \rm{x}}(-\beta \mathrm{E_{q}}^{+} )
+A_{q}^{+}\frac{\del^2}{\del^2 \rm{x}}(-\beta \mathrm{E_{q}}^{+} ) \nn  \\
&&+\left(\frac{\del^2 \Phi}{\del\mathrm{x}^2} 
  +\frac{\del^2 \Phi^*}{\del\mathrm{x}^2} \right) \mathrm{e}^{-\beta \mathrm{E_{q}}^{+}}\bigg\} \nn \\
&& +\frac{1}{g_{q}^{-}} \bigg\{C_{q}^{-} \Big[\frac{\del}{\del \rm{x}}(-\beta \mathrm{E_{q}}^{-} )\Big]^{2}
+\left(\frac{\del \Phi^*}{\del\mathrm{x}} \mathrm{e}^{-\beta \mathrm{E_{q}}^{-}} 
+ 4 \frac{\del \Phi}{\del\mathrm{x}} \mathrm{e}^{-2\beta \mathrm{E_{q}}^{-}}\right)\frac{\del}{\del \rm{x}}(-\beta \mathrm{E_{q}}^{-} )
+A_{q}^{-}\frac{\del^2}{\del^2 \rm{x}}(-\beta \mathrm{E_{q}}^{-} ) \nn \\
&& +\left(\frac{\del^2 \Phi^*}{\del\mathrm{x}^2} 
  +\frac{\del^2 \Phi}{\del\mathrm{x}^2} ) \right) \mathrm{e}^{-\beta \mathrm{E_{q}}^{-}}   
\bigg\} \bigg]
\eea

\begin{equation}
C_{q}^{+} = \Phi \mathrm{e}^{-\beta\mathrm{E_{q}}^{+}} +4 \Phi^* \mathrm{e}^{-2\beta\mathrm{E_{q}}^{+}} 
+3\mathrm{e}^{-3\beta\mathrm{E_{q}}^{+}}  
\end{equation}
\begin{equation}
C_{q}^{-} =\Phi^*\mathrm{e}^{-\beta \mathrm{E_{q}}^{-}} +4 \Phi \mathrm{e}^{-2\beta \mathrm{E_{q}}^{-}} 
+3\mathrm{e}^{-3\beta \mathrm{E_{q}}^{-}}  
\end{equation}







\end{document}